\documentclass{aa}  

\usepackage{graphicx}
\usepackage{epstopdf}
\usepackage{wasysym}
\usepackage{txfonts}
\newcommand{\msun}{\,$\rm M_{\odot}$} 
\newcommand{\rsun}{\,$\rm R_{\odot}$} 
\newcommand{\jinf}{$j_{\rm inf}$}

\begin{document}

   \title{Formation of massive close binaries}

   \subtitle{I. Detached evolution}

   \author{Mads S\o rensen
          \inst{1,2}
          \and
          Tassos Fragos\inst{2}
          \and
          Georges Meynet\inst{1}
          \and
          Lionel Haemmerlé\inst{1}}

   \institute{Observatoire de Genève, University of Geneva, Route de Sauverny, 1290 Versoix, Switzerland\\
              \email{Mads.Sorensen@unige.ch, Georges.Meynet@unige.ch, Lionel.Haemmerle@unige.ch}
         \and
             DARK, Niels Bohr Institute, University of Copenhage, Julia Maries vej 31, 2100 København Ø, Denmark\\
             \email{Anastasios.Fragkos@unige.ch}
             }

   \date{Received ; accepted }

 
  \abstract
   {The way massive pre-main sequence (pre-MS) binaries form and evolve is by large an open question. Here we systematically address this topic from the perspective of stellar structure and evolution models. We are looking at the pre-MS detached evolution of massive binaries that evolve, through accretion growth, from a small mass stellar cores to massive zero age main sequence binaries.}
   {We explore the initial conditions that lead to detached binary evolution of massive pre-MS binaries and ask how large a fraction of the observed binary systems may have been initially formed as low-mass protobinaries and later undergone a significant accretion phase while remaining detached.}
   {We develop a family of analytic models to describe the orbital separation, $a$, and mass ratio, $q$, evolution. For a given mass accretion rate onto the binary system, we define a recipe for distributing this mass between the two components. For this we 
   introduce a parameter $\eta$, such that $\tfrac{\dot{M}_2}{\dot{M}_1}=q^\eta$ at any time, to determine the binary mass ratio evolution. Depending on the choice of $\eta$, any type of mass ratio evolution is possible. Furthermore, we use MESA, a detailed stellar structure code, to calculate an extensive grid of binary sequences where a protobinary undergoes accretion, and we identify the initial conditions that separates detached from non-detached pre-MS binary evolution.}
   {A value of $\eta$ around 2 allows accretion growth in detached systems to form close massive binaries on the Zero Age Main-Sequence with minimum orbital periods down to about 1.2 days for $M_{\rm 1,ZAMS}=20-30$\msun{} twin-binaries. $\eta=2$ can also reproduce the observed population of binary systems with primary stars above 6\msun.
   }
   {
   The whole observed range of massive close binaries can form via accretion growth in detached systems, making the binary formation channel of accretion growth a strong contender to explain the formation of massive close binaries, including progenitors of coalescing binary black holes.}


   \maketitle
%

\section{Introduction}\label{sec:intro}
The formation of massive stars, >8\msun{}, is by large an open question, and even more so, is the formation of massive binary stars. This is in part because massive stars are rare and short lived, but also because their formation process lies hidden inside optical thick natal clouds that are not removed until the massive stars launch strong winds or some of them undergo supernovae whereby the cloud is cleared. However, this only happens after the massive stars have formed.

We would expect the formation of massive single and binary stars to be subjected to the same physical processes and environments. Therefore we begin by a brief discussion on massive single star formation. There are two proposed ways to form massive single stars, both of which involves lower mass protostars. These are the accretion and the merging scenarios, respectively. In the accretion scenario a less massive star can accrete material from its surroundings and grow to become massive. The alternative is the merging scenario where the merging of two intermediate mass stars can produce a single massive star.

How massive a star can become via accretion is determined from its metallicity and the geometry of the infalling material \citep[e.g.][]{larson1971, kuiper2010}. The stellar structure of single stars undergoing accretion has been studied by many authors \citep[e.g.][]{palla1992, behrend2001, omukai2001, omukai2003, yorke2008, hosokawa2010, hosokawa2013, hosokawa2016,lee2016, haemmerle2016, haemmerle2017}. Yet, open questions remain in this picture of which we note challenges on relations between the accretion rate, metallicity and swelling due to internal restructuring and shell D-burning, and transport of angular momentum and its impact on stellar rotation. How is angular momentum from the infalling material coupled to the star and transported with the accretion flow into the star and what is the effect on the stellar structure?
A potentially important effect is the modified Eddington effect due to rapid rotation, also known as the $\Omega\Gamma$-limit \citep{maeder2000}. A rapid rotating star has a lower effective gravity due to the centrifugal force and this gives rise to a lower modified (effective) Eddington luminosity limit. In accreting stars angular momentum is carried into the star which is then spun up, and may reach a critical rotation, thus stopping the accretion. \citet{lee2016} looked at the $\Omega\Gamma$-limit in the formation of massive population III stars via accretion growth. They concluded that protostars quickly reaches a critical Eddington limit, whereby the accretion stalls and as a result producing stars more massive than the 20-40\msun{} seems impossible due to the $\Omega\Gamma$-limit. \citet{lee2016} treated rotation of their models in post processing and assumed solid body rotation. 
\citet{haemmerle2017} used a new version of the Geneva Stellar Evolution code to follow the accretion growth of protostar, keeping track of its structure and rotation profile.
This allowed \citet{haemmerle2017} to observe how differential rotation plays a crucial role in preventing the star from reaching the $\Omega\Gamma$-limit already at the swelling phase postulated by \citet{lee2016}.
As is shown in \citet{haemmerle2017} the formation of stars more massive than 40\msun{} is possible, if the protostar can dissipate 2/3 or more of the angular momentum gained from the accretion. This is possible by a strong breaking mechanism which modulate the angular momentum accretion history of the star.

The second possibility, is to grow low mass stars into high mass stars through merger events, which requires a high stellar number density \citep{bonnell2001,bonnell2005}.
The stellar structure resulting from the merger of two stars was investigated by \citet{glebbeek2013}. They found the mass loss in the merger event for head-on collisions is less than 10 \%, which permit the production of massive stars via mergers. \citet{schneider2016} argued the merger event produces large scale magnetic fields in the new star, and if so, that about 10 \% of OBA pre-MS and MS which displays such magnetic fields are merger products.
\citet{baumgardt2011} performed N-body simulation of young star forming embedded clusters, accounting for the process for accretion growth onto each star. They find, that increasing the stellar number density of the embedded cluster increases the number of interactions. However, in many cases, the close encounter does not produce a merger, but a runaway star instead. These conclusions were revised by \citet{railton2014} who used pre-MS isochrone tracks of stars from 1-8 \msun{} to describe the earliest stages of stellar evolution. They find that collisions were common and could produce high mass stars. However, using pre-MS isochrone tracks of constant mass ({\it i.e.} non-accreting) stars gives an unrealistically large cross-section for interactions. 



Observing massive stars in the making is very hard as they are surrounded by infalling material and outflows. If they are sitting in an embedded cluster, still surrounded by their natal cloud, observations are even harder. 
Since stars, including binaries, are generally believed to form inside embedded clusters \citep{lada2003}, both accretion and dynamical interactions should play a role in forming binary systems: accretion, because the embedded cluster offers a reservoir of gas, and dynamical interactions due to the potentially high stellar number density. This naturally leads to ask if massive binary stars may also form from low-mass proto binaries and accretion growth without reaching the Roche limit or being scattered. 

An interesting example is the observation of the HII outflow region S106, which contains a massive pre-MS binary, S106IR, with an orbital period of 5 days, primary mass of $\approx19$ \msun{} and mass ratio $q\approx0.17$ \citep{comeron2018}. The sparse number of other sources in close vicinity makes S106IR a prime candidate for massive binary formation via accretion growth.

Another candidate is IRAS 04191+1523 which is a very low mass binary.
It has a high mass ratio $q\approx0.85$ and orbital separation $a\approx860$ au \citep{lee2017}. \citet{lee2017} proposed that IRAS 04191+1523 formed from a process of fragmentation in a turbulent medium, as initially suggested by \citet{bonnell2005}.

Massive binaries can result from dynamical interactions, when, e.g., two massive stars form a bound system as they pass near each other at sufficiently low velocities for their mutual gravitational attraction to govern their further dynamics. If dynamic interactions would be the means by which the majority of binary stars form, then 
the stellar structure due to accretion growth is not relevant. We think however that although dynamical interactions can be important, especially in dense stellar systems, it is probably not the only channel through which close massive binary stars are formed.

Here we focus on the formation of binaries via accretion growth and we define a binary system to be composed of a primary star $M_1$ and a companion star $M_2$ separated by some orbital separation $a$ and eccentricity $e$, with an orbital period $P$. We further define the binary mass ration $q\equiv \tfrac{M_2}{M_1}$ and call a binary with $q\geq0.95$ a twin-binary \citep{lucy1979}.
The accretion process requires that a low-mass protobinary is concepted out of the gas phase. Three theories has been proposed to describe the formation of protobinaries in molecular clouds. The oldest one is the fission theory, where a rapidly spinning proto-star splits into two \citep{jeans1919}. However, the compressible fluids from which stars form, work against this formation scenario, as the rapidly spinning protostar generates spiral arms that dissipate angular momentum of the protostar before it can be split \citep{tohline2002, bate2015}.
Alternatively, a protobinary forms from the gravitational fragments of a collapsing cloud, due to local density fluctuations which produce areas of smaller free fall times, hence a faster collapse. These fragments are then bound once they fragment and start their contraction \citep{boss1979, boss1988}.
Finally, it has been proposed that sufficiently cool and massive protostar discs may be dynamically gravitational unstable and fragment to form a protobinary \citep{kratter2006}. In the present paper, we will remain agnostic about the exact protobinary formation scenario and we will assume that protobinaries are formed at relative low mass and undergo a phase of accretion growth that brings them to their final ZAMS masses. During this phase we ignore any potential dynamical interactions with the rest of the cluster or merger events.

Several other studies of protobinaries undergoing accretion growth have concentrated on the accretion of material onto a protobinary of some mass ratio using hydrodynamic simulations \citep[][]{bate1997ballistic, bate1997gas,ochi2005,young2015a,young2015b}, but little is known on their stellar structure evolution and its role in binary formation. 
To our knowledge, the earliest study of accreting binaries was by \citet{rayburn1976} who considered accretion onto a main sequence binary that encounters a interstellar cloud. \citet{artymowicz1983} studied for the first time the pre-main sequence binary stars undergoing accretion growth. He used a three-body approximation to describe the accretion onto each star. The first to include the stellar structure to study the formation of binaries was \citet{tutukov1983}. He noticed that the orbital period distribution featured a distribution with a clear break around orbital periods of 100 yr, indicative of two formation mechanisms. \citet{tutukov1983} assumed binary formation via the fission process and that wide binaries ($P_{\rm orb} \gtrsim 100$ yr) fissioned out the collapsing protocloud before a hydrostatic-equilibrium gas-dust core could form. The close binaries ($P_{\rm orb}\lesssim 100$ yr) would fission out of the protocloud after the gas-dust core has formed.
The most recent study on the stellar structure of accreting proto-binaries, to our knowledge, is by \citet{krumholz2007}. They used a simple one-zone model by \citet{mckee2003}, under the assumption of constant accretion rates, for binaries in circular orbits, and with fixed mass ratio. \citet{krumholz2007} conclude that in case that mass transfer occurs, it is always the primary star that fills its Roche lobe and this is due to internal restructuring and subsequent shell D-burning. They also find that mass transfer is unstable unless the initial mass ratio $q \geq 0.8$ and the donor is in the D-shell burning. Overall they suggest that mass transfer can explain the observed excess of twin-binaries.

Many open questions on the formation via accretion remains to be highlighted. There is no reason to suspect, that the binary mass ratio remains constant during the accretion phase. It is not clear how close of a massive binary can be formed from accretion, as this depends both on the initial orbital separation- and the mass ratio-evolution, as well as the protobinary's accretion history. We expect the orbital period- and mass ratio-evolution to also put limits on the fraction of the binary population that can be explained as originating through an accretion process. Finally, it is not clear that the primary star is the only star that can overflow its Roche lobe during the pre-MS accretion phase. In fact, we will show that even under the assumption of constant mass ratio, the secondary star may also overflows its Roche lobe while undergoing accretion.

In this study, we will simulate the accretion phase of a pre-MS binary system using the 1D stellar evolution code Modules for Experiements in Stellar Astrophysics \citep[MESA][]{paxton2011,paxton2013,paxton2015,paxton2018} to evolve, in detail, the stellar structure of each star as they accrete. Further, the MESA binary module allows us to track the orbital separation as the stars accrete. Our main goals in this first paper are to study the detached evolution of binaries and investigate the relation between initial binary orbit with post-accretion binary orbits, primary mass and mass ratio. The pre-MS evolution of binary stars undergoing accretion is a complex study and we address here only detached evolution in circular orbits. Effects of mass transfer, eccentricity, and tidal evolution during the pre-MS of accreting binary stars is beyond the scope of the present paper.
Here we develop an analytic model for a pre-MS binary system undergoing accretion, that account for the evolution of orbital separation $a$ and mass ratio $q$. Further, we simulate a large grid of binary accretion sequences generating binaries with primary star masses between 6-60\msun{} at zero age main sequence (ZAMS) given different initial conditions. From the grid we deduce the limit separating detached from non-detached binary formation via accretion growth at the ZAMS. This detached vs. non-detached limit at ZAMS, we compare to the observed ZAMS binary populations mass ratio- and orbital period distribution for the relevant range of primary masses, to see if any observed binary systems at ZAMS suggests a non-detached pre-MS evolution.

Our paper is built up as follow. In sect. \ref{sec:theory} we develop our analytic model of pre-MS binaries undergoing accretion. First, we consider in sect. \ref{sec:orbit_evo}, the orbital separation evolution dependent on the change in pre-MS binaries' orbital angular momentum as the protostars accrete material from some external source. Secondly, in sect \ref{sec:mass_ratio_evo}, we present our model for the change in mass ratio in response to the accretion. In particular we introduce a new parameter $\eta$ that governs the change in mass ratio. In Sect. \ref{sec:eta} we briefly review studies using hydrodynamic simulations of protobinaries undergoing accretion and estimate $\eta$ from these sources. We also present analytic estimates of $\eta$.
Section \ref{sec:single_star} describes our MESA model of an accreting pre-MS star and discusses its structure and evolution while accreting. Section \ref{sec:grids} presents our grids of pre-MS binary accretion sequences. Sections \ref{sec:result}, \ref{sec:discussion}, and \ref{sec:conclusion} are reserved to present our results, discussion and conclusion respectively.

\section{Accretion growth of pre-main sequence binaries}\label{sec:theory}
In general, we will not be concerned with the exact process from which a protobinary is formed, fragmentation or capture, however, we will explore the case where a low-mass protobinary is undergoing accretion growth to reach its final ZAMS mass. A consequence of two protostars forming at the same time is that they will have near-equal jeans masses. Thus, assuming the two protostars  becomes bound, this scenario yields an initial protobinary mass-ratio distribution biased to high mass-ratios. Subsequent accretion growth can then bring the protobinary close and change the mass ratio.

\subsection{Orbital evolution of an accreting protobinary}\label{sec:orbit_evo}
To understand the effect of accretion on to a protobinary, we consider a circular orbiting point mass model with masses $M_1$ and $M_2$ at an orbital separation $a$. The protobinary then has an orbital angular momentum

\begin{equation}\label{eq:Jorb}
    J_{orb}^2 = \frac{M_1^2M_2^2}{M}Ga
\end{equation}
where $G$ is the Newtonian gravitational constant and $M = M_1 + M_2$ is the total mass of the protobinary. We now imagine that the protobinary is accreting infalling gas from a gas reservoir external to the system. Let $\dot{M}$, $\dot{M_1}$, and $\dot{M_2}$ be the accretion rate of the protobinary, the primary and secondary protostar respectively. The effect of accretion to the protobinary's orbital separation depends on the change of its binary component masses and its orbital angular momentum which is found from differentiating eq. \eqref{eq:Jorb} with respect to time. This gives
\begin{equation}\label{eq:dadt}
    \frac{\dot{a}}{a} = 2 \frac{\dot{J}_{\rm orb}}{J_{\rm orb}} - 2\frac{\dot{M}_1}{M_1} - 2\frac{\dot{M}_2}{M_2} +\frac{\dot{M}}{M}
\end{equation}
where a dot above a letter denotes first order differentiation with respect to time. As the protobinary system is undergoing accretion the first term on the right hand side takes the form
\begin{equation}\label{eq:dotJ/J}
    \frac{\dot{J}_{\rm orb}}{J_{\rm orb}} = \frac{\dot{J}_{\rm acc}}{J_{\rm orb}}
\end{equation}

\subsubsection{Spherical symmetric accretion}
If the infalling gas is accreted spherically symmetricly on to each protostar, the gas contributes with zero net angular momentum to the system, i.e. $\dot{J}_{acc} = 0$, and the orbital angular momentum is constant. Solving eq. \eqref{eq:dadt} for spherically symmetric accretion gives


\begin{equation}\label{eq:ai2af_spherical}
    \frac{a_f}{a_i} = \left(\frac{M_{i}}{M_{f}}\right)^3\frac{(1+q_f)^4}{q_f^2}\frac{q_i^2}{(1+q_i)^4}
\end{equation}
where subscripts $i$ and $f$ are initial and final state of the system and $q$ is the binary mass ratio. Thus the orbit shrinks proportionally to the ratio of the initial to final mass of the binary cubed, and is less dependent on the change of initial to final mass ratio. 

\subsubsection{Disc accretion}
Another possibility is that the infalling gas settles into a system of discs around the protobinary before being accreted. Such a system of discs could consist of first a circumbinary disc that gradually captures gas from the external cloud. Once the gas has settled into the circumbinary disc it wanders inwards before eventually falling towards either of the two protostars where the gas is captured into a protostellar disc that surrounds each protostar. If a circumbinary disc is necessary is not known and we will ignore it for now. Gradually the material is transported through the protostellar disc before being accreted onto the protostar. In the gradual descent from far away and down to the protostar, the infalling gas looses any excess angular momentum such that at the point of accretion, it has specific angular momentum equal to the specific angular momentum of each accreting protostar, $j_{1,2}=r_{1,2}^2\omega$, i.e. $\dot{J}_{\rm acc} = \dot{M}_1r_1^2\omega+\dot{M}_2r_2^2\omega$ . Equation \eqref{eq:dotJ/J} then becomes
\begin{equation}
    \frac{\dot{J}_{orb}}{J_{orb}} = \frac{\dot{M}_1r_1^2\omega+\dot{M}_2r_2^2\omega}{J_{orb}}
\end{equation}
and eq. \eqref{eq:dadt} has the analytic solution
\begin{equation}\label{eq:afai_disk}
    \frac{a_f}{a_i} = \frac{M_i}{M_f}
\end{equation}
which is independent on the associated change in mass ratio. The corresponding solution in orbital period is
\begin{equation}\label{eq:pfpi_disk}
    \frac{P_f}{P_i} = \left(\frac{M_i}{M_f}\right)^2 \rm .
\end{equation}

It is clear that relative to spherically symmetric accretion, disc accretion shrink the orbit at a slower rate. Hence, accretion via a disc allows the largest increase in mass of the protobinary before initiating RLOF and thus the protobinary can avoid merging.
This is expected since spherically symmetric accretion does not add angular momentum to the system, only mass is being added. Hence, as the stellar mass increases, the orbits needs to shrink significantly to conserve the angular momentum. During disc accretion the stars receive some angular momentum allowing them to stay detached for a greater mass increase.

Here we have ignored eccentric orbits when we deduced the orbital separation change in response to accretion onto each star. However, it is unclear, that accretion of material from an external source would keep the binary orbit from becoming eccentric. I.e. if the specific angular momentum of the accreting material is larger than that of the accretor, potentially the orbit would become eccentric. We note that disc accretion resembles the orbital evolution of isotropic wind mass loss in massive stars as described by \citet{dosopoulou2016}. They find that the period averaged rate of change in the orbit's eccentricity is $\langle \dot{e}\rangle = 0$. Another interesting point about accretion via a protostellar disc is the radiation-problem of stars undergoing accretion, as discussed in sect. \ref{sec:intro}. Disc accretion actually allows the protobinary to overcome the radiation-barrier and adds small amounts of angular momentum to keep the two protostars separated. We will thus assume that protobinary stars undergoing accretion growth receives material with a specific orbital angular momentum equal to the accretor. Further we assume circular orbits through the entire accretion period and we do not consider effects of stellar rotation.

\subsection{The accretion-growth mass ratio evolution of a protobinary}\label{sec:mass_ratio_evo}
As material is accreted onto each protostar the infalling material is being split into two parts such that $\dot{M} = \dot{M}_1 + \dot{M}_2$. Hence, $\dot{M}_1 = f_1\dot{M}$, $\dot{M}_2=f_2\dot{M}$, and we have the constraint that $f_1+f_2=1$. The two factors $f_1$ and $f_2$ thus describe the fraction of the total accretion rate that is given to each protostar.
We can intuitively expect that the infalling material is split based on some dependence on the mass ratio at any given time. Here we assume a dependence of the form:
\begin{equation}\label{eq:qeta}
    \frac{\dot{M}_2}{\dot{M}_1} = q^{\eta} \, ,
\end{equation}
where $q \equiv M_2/M_1$, and $\eta$ dictates the splitting of material between the two protostars. A similar approach was assumed by \citet{artymowicz1983}. We will discuss a physical understanding of $\eta$ in Sect. \ref{sec:eta} but for now just assume it is a constant. Eq. \eqref{eq:qeta} can be written as
\begin{equation}
\frac{\dot{M}_2}{M_2}\frac{M_1}{\dot{M}_1} = q^{\eta-1}
\end{equation}
hence the relative increase of $M_2$ divided by the relative increase of $M_1$ varies as $q^{\eta-1}$. 
We see that if $\eta$=1, the relative increase remains the same at every time and thus the initial $q$ is also the final one.
If $\eta<1$, and since initially $q<1$, the relative increase of $M_2$ is larger than the relative increase of $M_1$. So $M_2$, which initially smaller than $M_1$, will catch up $M_1$ and equalize the two masses. Thus the ratio converges towards 1. The smaller $\eta$ is, the more rapid the convergence of the binary mass ratio to 1. If $\eta>$ 1, the relative increase of $M_2$ is always smaller than the relative increase of $M_1$. $M_2$ will thus never catch up $M_1$ and the mass ratio will tend towards 0. The greater $\eta$ is the faster the convergence to 0 is. When the masses are equal, i.e. $q=1$ there is no change in mass ratio. 

\begin{figure}
    \centering
    \includegraphics[width=1.0\linewidth]{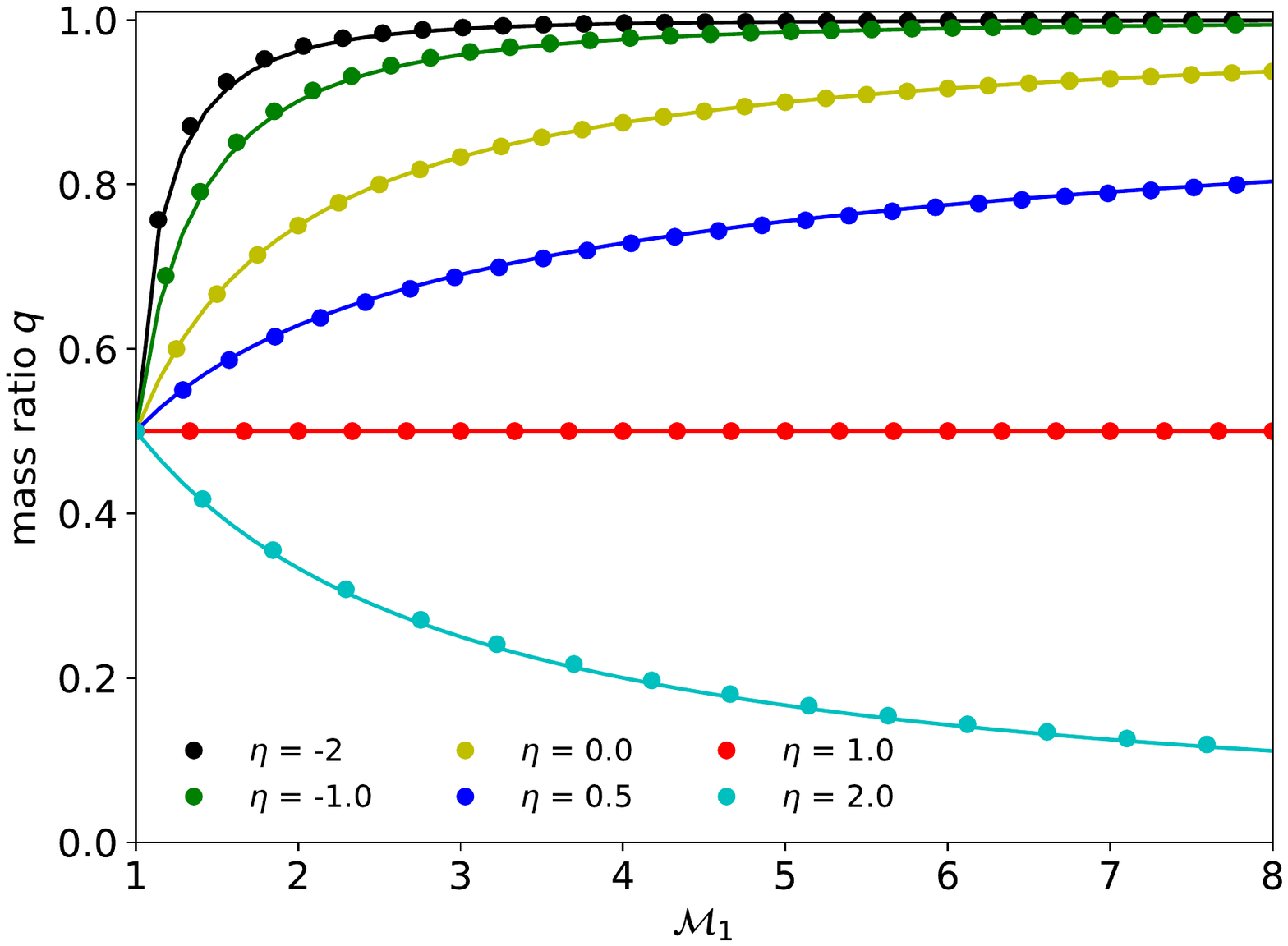}
    \caption{The evolution of a protobinary's mass ratio $q$ as a function of the primary mass increase factor $\mathcal{M}_1$, assuming different values of $\eta=[-2, -1, 0, 0.5, 1, 2]$ (black, green, yellow, blue, red, and light blue respectively). The initial mass ratio is always $q_i=0.5$.
    When $\eta = 1$, $q$ remains constant, if $\eta<1$, $q$ goes to $1$, and if $\eta>1$ $q$ evolves towards $0$.}
    \label{fig:qi2qf}
\end{figure}

From the definition of $q$, we can evaluate the evolution of the mass ratio by simple differentiation with respect to time and obtain
\begin{equation}\label{eq:dqdt}
    \dot{q} = (q^\eta-q)\frac{\dot{M_1}}{M_1} \rm .
\end{equation}
We again see, that if $\eta = 1$, $\dot{q}=0$ and the mass ratio stays constant as the protobinary is accreting. If $\eta \neq 1$ we can integrate eq. \eqref{eq:dqdt} and get the following solution
\begin{equation}
    \frac{|q_f^{1-\eta}-1|}{|q_i^{1-\eta}-1|} = \left(\frac{M_{1,f}}{M_{1,i}}\right)^{\eta-1} \, .
\end{equation}
Rearranging this expression yields reversible relations between initial and final mass ratio with a dependence on the initial and final mass of the primary protostar $M_{1}$ and $\eta$.
First is the initial to final mass ratio relation
\begin{equation}\label{eq:qi2qf}
    q_f  = \left[\left(\frac{M_{1,f}}{M_{1,i}}\right)^{\eta-1}|q_i^{1-\eta}-1|+1\right]^{\frac{1}{1-\eta}}
\end{equation}
and second the reverse solution
\begin{equation}\label{eq:qf2qi}
    q_i  = \left[\left(\frac{M_{1,i}}{M_{1,f}}\right)^{\eta-1}|q_f^{1-\eta}-1|+1\right]^{\frac{1}{1-\eta}} \, .
\end{equation}
The ratio $\frac{M_{1,i}}{M_{1,f}}$ suggests the mass ratio is sensitive to the times that the primary protostar mass is increased during accretion. \citet{artymowicz1983} defined the ratio $\mathcal{M} = \frac{M_f}{M_i}$ as the total mass increase factor.
Using the total mass increase factor and remembering that $M_1 = \frac{M}{1+q}$, rearranging eq. \eqref{eq:dqdt} gives 
\begin{equation}
    \mathcal{M} = \left(\frac{|q_f^{1-\eta}-1|}{|q_i^{1-\eta}-1|}\right)^{\frac{1}{\eta-1}}\left(\frac{1+q_f}{1+q_i}\right) \text{ ,}
\end{equation}
which is the solution found by \citet{artymowicz1983}. Here, we will however use the primary mass increase factor $\mathcal{M}_1 = \frac{M_{1,f}}{M_{1,i}}$. 
In Fig. \ref{fig:qi2qf} we show the numerical (dots) and analytic (solid lines) solution for a protobinary's mass ratio evolution as a function of the primary mass increase factor, $\mathcal{M}_1$, for different values of $\eta$. The numerical values are found from solving numerically equations \eqref{eq:qeta} and the constraint $\dot{M}_1+\dot{M}_2 = \dot{M}$. The initial conditions and $\dot{M}$ are not important since it is the relative increase of $M_1$ over its initial value that is shown along the x-axis. If $\eta=1$ (red color) the mass ratio stays constant. If $\eta<1$ (blue yellow, green, and black colors) accretion drives the mass ratio towards unity. Making $\eta$ smaller, the rate with which, the mass ratio asymptotically converges to unity increases. Hence, for extreme values of $\eta$, i.e. $\eta<<1$, then the accretion rate of the secondary protostar will initially be $\dot{M}_2\sim \dot{M}$ and quickly the mass ratio will tend to $q\sim1$, at which point the two stars accrete at equal rates, i.e. $\dot{M}_2\sim\dot{M}_1$ and $f_1\sim f_2\sim\tfrac{1}{2}$. So effectively, as unity is reached, the protobinary splits the infalling gas in equal parts. If $\eta>1$ (light blue color) accretion drives the mass ratio towards 0. Again, the larger $\eta$ is, the stronger the asymptotic convergence towards 0 is.

Three interesting result come out of eq. \eqref{eq:qi2qf}. First is the evolution of $q$ when $0<\eta<1$. In this regime, the primary protostar receives the largest fraction of accreted material, yet, the mass ratio is asymptotically increasing towards unity, which means that $\dot{M_1}/M_1<\dot{M_2}/M_2$ but $\dot{M_1}>\dot{M_2}$.

\begin{figure}
    \centering
    \includegraphics[width=1.0\linewidth]{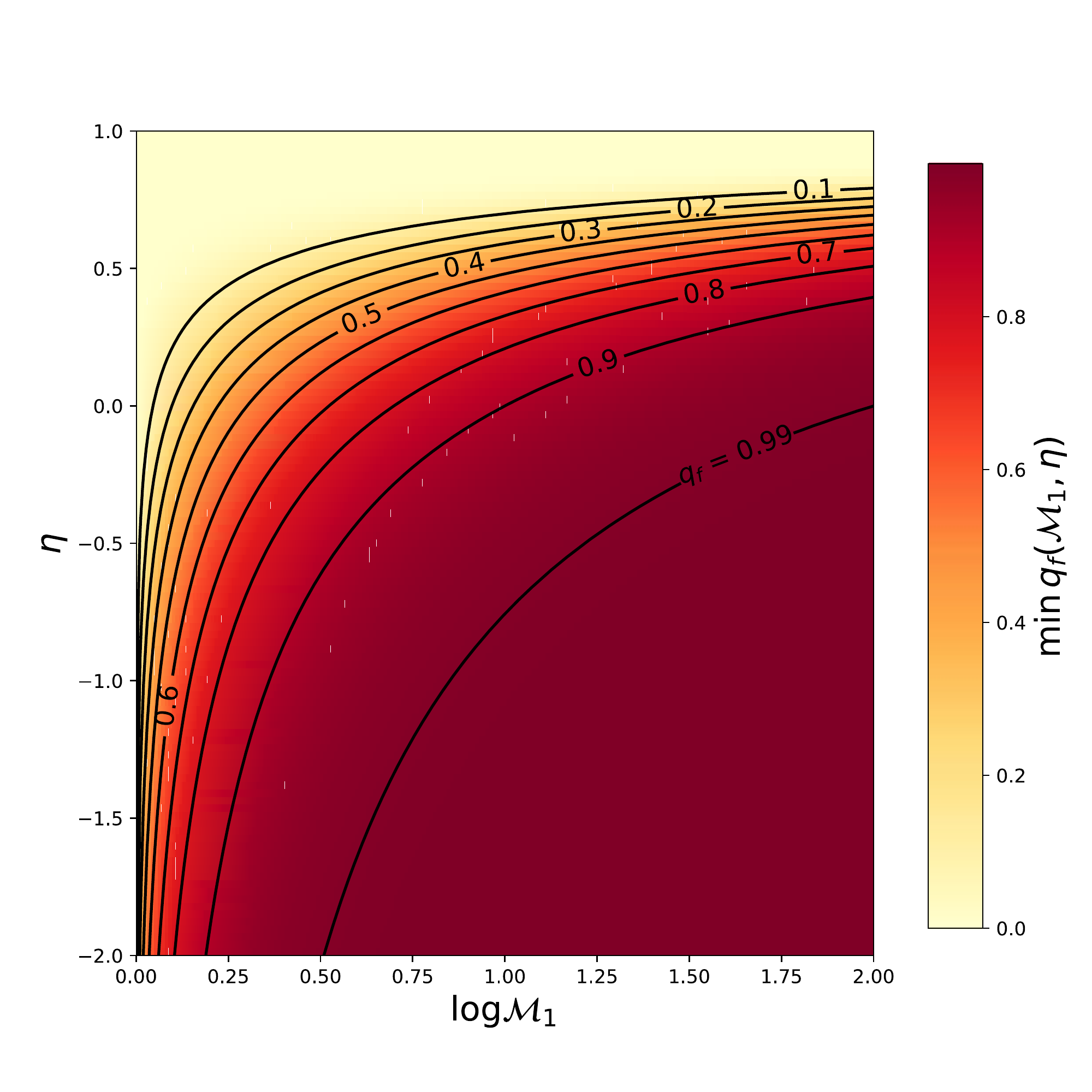}
    \caption{Minimum final mass ratio, $\min q_{f}$, as a function of the primary mass increase factor $\mathcal{M}_1$ and $\eta$, eq. $\eqref{eq:limqf}$. The solid black lines are levels of constant final binary mass ratio from $q_f=0.1$ to $q_f=0.99$.}
    \label{fig:eta_qi2qf}
\end{figure}

The second interesting result of eq. \eqref{eq:qi2qf} is the limit 
\begin{equation}\label{eq:limqf}
\lim_{q_i \rightarrow 0} q_f(\mathcal{M}_1,\eta) = \left[-\mathcal{M}_1^{\eta-1}+1\right]^{\frac{1}{1-\eta}} \, ,
\end{equation}
when $\eta<1$. 
In Fig. \ref{fig:eta_qi2qf} we show the minimum $q_f$ that is reached for all $\eta$'s in the range [1.0; -2.0] as a function of the primary mass increase factor. This figures tells us that a system starting with a primary mass of 0.5\msun{} and increasing it 10 times to 5\msun{}, then for all $\eta$ inferior to about -0.8 one obtains a nearly equal mass ratio. Even if the initial mass ratio is very small, i.e even for very low mass secondary stars. If the final mass of the primary is greater (typically 50 M$_\odot$), the final mass ratio is nearly 1 for all $\eta$ inferior to about 0.0. In that case, to obtain mass ratios smaller than about 0.1, $\eta$ larger than about 0.7-0.8 should be considered. Mass ratios in between are reached for a rather restricted domain of $\eta$, between 0.0 and 0.8. This tells us that starting from a very small mass ratio, a massive star primary will either produce again a very small ratio or a mass ratio equal to 1 for most of the values of $\eta$. Only in a relatively modest interval of $\eta$ other mass ratios can be reached. The third interesting fact to notice about the binary mass ratio parameterisation via $\eta$ is that it allows practically for any type of binary mass ratio evolution.



\subsection{The value of $\eta$?}\label{sec:eta}
In the previous section we studied qualitatively the effect of varying $\eta$ without associating its value to the underlying physical processes that govern it.

In this section we try to estimate values of $\eta$ based on different assumptions: i) a scenario of independent competitive accretion, where the protobinary and two protostars undergo Bondi-Hoyle accretion with the accretion radius and accretion rate being dependent firstly on the sound speed and secondly on the binary components' orbital velocities, ii) independent accretion following some mass-luminosity relation, iii) a close binary where the effective accretion radius is given by the the Roche radius of each protostar. Finally we look at relevant hydrodynamic simulations to estimate $\eta$ from these and compare with our estimates.

\subsubsection{Bondi-Hoyle competitive accretion}\label{sec:BHaccretion}
In the scenario of competitive accretion, the splitting of infalling gas, hence $\eta$, depends on the accretion radius of each protostar 
\begin{equation}\label{eq:Racc_BH}
    R_{acc} = \frac{2GM_{\ast}}{(c_s + v_{\ast} )^2} \, ;
\end{equation}
here $M_{\ast}$ is the mass of the protostar, $c_s$ is the sound speed of the infalling gas, and $v_{\ast}$ is the streaming velocity of the gas relative to the protostar. The mass accretion rate of the protostar is
\begin{equation}\label{eq:dmdt_BH}
    \dot{M}_{\ast} \sim \pi R_{acc}^2\rho (c_s+v_{\ast})
\end{equation}
where $\rho$ is the density of the in-falling gas far away from the source.

The first limiting case is to assume that the streaming velocity of the in-falling material relative to the protostar is negligible, and only the sound speed is relevant for the gravitational radius of each protostar.
Assuming each protostar has the accretion rate given by eq. \eqref{eq:dmdt_BH} with $v_{\ast}<< c_s$ (this could be the case of a wide binary system), the accretion rate of the secondary- to the primary protostar, can be equated with eq. \eqref{eq:qeta} to yield
\begin{equation}
    \frac{\dot{M}_2}{\dot{M}_1} = \left(\frac{M_2}{M_1}\right)^{2} = q^{2}
\end{equation}
hence $\eta = 2$. According to the analytic analysis made above, this favors systems where the final mass ratio evolves towards small values.

The second limiting case, is to assume the sound speed is negligible relative to the streaming velocity of the in-falling gas relative to each protostar, i.e. $v_{\ast}>>c_s$. In this case we assume that the relative streaming velocity is equal to each protostar's orbital velocity
\begin{equation}\label{eq:vel_orbit}
    v_{1,2} = \omega r_{1,2} = \sqrt{\frac{GM}{a^3}}r_{1,2}
\end{equation}
For a circular orbit the relative position of each protostar is
\begin{equation}\label{eq:r_pos}
    r_1 = -\frac{M_2}{M}a \\ r_2 = \frac{M_1}{M}a \rm .
\end{equation}
Combining eqs. \eqref{eq:Racc_BH},\eqref{eq:dmdt_BH}, \eqref{eq:vel_orbit}, and \eqref{eq:r_pos} with eq. \eqref{eq:qeta} we see that
\begin{equation}
     \frac{\dot{M}_2}{\dot{M}_1} = \left(\frac{M_2}{M_1}\right)^{5} = q^5
\end{equation}
and hence $\eta = 5$. This would suggest that the accretion onto a binary system is strongly pushing the mass ratio towards 0, hence, the primary protostar is effectively accreting all the infalling mass.

\subsubsection{Relative accretion rate from Roche spheres}
Rather than the accretion radius dictating the relative accretion rates, the Roche lobe radii could control this. The estimate of $\eta$ under this assumption was in fact done by \citet{rayburn1976}. He did this by solving the restricted three body problem of a binary system following the trajectory of test particles. Specifically he calculated the trajectory of several test particles at different initial positions far away from the binary system and followed them down towards the binary system. His computation of the individual test particles ended when they hit the Roche lobe of either of the two stars. Under the assumption that the test particle was accreted once it hits the Roche lobe, he calculated the relative rate of accretion onto each star, which is shown in his figure 1. In our terminology he estimated $\eta=3$ if the mass ratio $q_i>0.3$.

The second estimate given in Sect. \ref{sec:BHaccretion}, where the orbital velocity of each proto-star governs the relative accretion rate, ignores the presence of the Roche potential. However, for close binaries, this will govern the value of $\eta$, rather than the orbital velocities. Hence, $\eta=3$ seems more realistic over $\eta=5$. 


\subsubsection{Independent accretion}
Assume now a protobinary where the orbital separation is so large that the two stars do not interact with each other as they accrete.
The value of $\eta$ in such a scenario is possible to estimate from two ways. In the first way we can assume that each star is accreting with the same accretion rate $\dot{M'}$, in which case
\begin{equation}
    \frac{\dot{M}_2}{\dot{M}_1} = \frac{\dot{M'}}{\dot{M'}} = 1 \,,
\end{equation}
which from our parameterisation given by eq. \eqref{eq:qeta} has the solution $\eta=0$ which means that the accretion rate onto each star is independent on the binary mass ratio $q$. If $q=1$ the value of $\eta$ is irrelevant.

A second possibility is a parameterised accretion rate either from observations \citep[e.g. ][]{behrend2001} or hydrodynamic simulations \citep[e.g. ][]{haemmerlepeters2016}. Here we use the Churchwell-Henning accretion (CH) law put forth by \citet[][see also Sect. \ref{sec:CH_law}]{behrend2001}.
We have computed an accretion sequence of a single star using the CH law which is shown in Fig. \ref{fig:eps_nuc}. To estimate a value of $\eta$ from the CH law we take the accretion rate of the star in the final time step of the model, i.e. when the mass of the star is 60\msun{}, as $\dot{M}_1$. As $\dot{M}_2$ we take all accretion rates when the same star is less massive than 60\msun{}. This produces a range of mass ratios from $q = [\frac{1}{60},1]$. For the binary mass ratios $q<0.5$ the behaviour of the ratio $\dot{M}_2$/$\dot{M}_1$ is affected by the smaller star first contracting and then restructuring internally, see also sect. \ref{sec:single_star}, but in the mass ratio interval $0.5 < q <1$, the relative accretion rate of the two stars is such that the mass ratio remains constant, hence $\eta=1$. The behaviour for $q<0.5$ is not significant different from $\eta=1$ so it is reasonable to assume that independent accretion has value of $\eta=1$.

\subsection{$\eta$ from hydrodynamic simulations of accreting proto-binaries}
\begin{figure}
    \centering
    \includegraphics[width=1.0\linewidth]{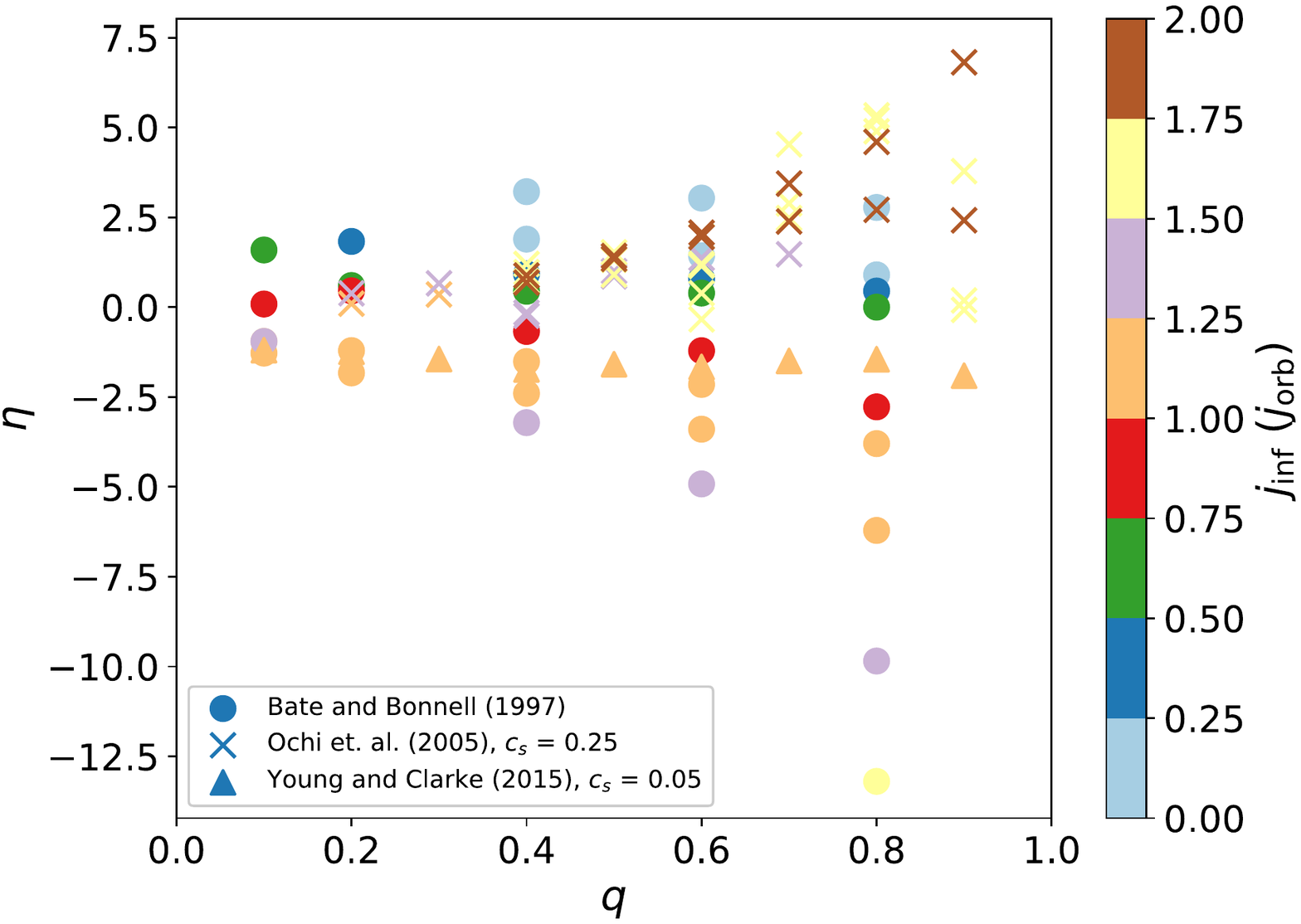}
    \caption{Estimated values of $\eta$ as a function of binary mass ratio $q$ from hydrodynamic simulations by \citet{bate1997gas} (circles), models with a sound speed $c_s=0.25$ by \citet{ochi2005} (crosses), and \citet{young2015b} with sound speed $c_s$ = 0.05 (triangles). The sound speed is normalized relative to the orbital velocity of the binary. The color indicates the initial specific angular momentum of the accreted material. The way we obtained the data points is given in appendix \ref{appendix:eta}. 
    }
    \label{fig:eta_q_estimates}
\end{figure}

The precise value of $\eta$ is unknown and determining it, is not a simple enterprise. Besides our estimates in sect. \ref{sec:BHaccretion}, we also look at studies calculating $\eta$ numerically, and find strong dependencies on the initial conditions of the in-falling material, the binary mass ratio, and even the details of the numerical method used. This can be seen in Fig. \ref{fig:eta_q_estimates} which shows different estimates of $\eta$ from studies of hydrodynamic simulations of protobinaries undergoing accretion. Different values of $\eta$ are shown as a function of the mass ratio $q$. The color indicates the initial specific orbital angular momentum of the accreted material in units of orbital angular momentum. The legend refers to the three studies included in the figure. Dots are from \citet{bate1997gas}, crosses are from \citet{ochi2005} with sound speeds of $c_s=0.25$ in units of orbital velocity, and triangles are from \citet{young2015b} cold models with $c_s=0.05$ in units of orbital velocities.

In two papers, \citet{bate1997ballistic} and \citet{bate1997gas} looked at the mass ratio rate of change for different values of the specific orbital angular momentum $j_{\inf}$
\begin{equation}\label{eq:jorb_specific}
    j_{\rm inf} = \alpha j_{orb} = \alpha \frac{q}{(1+q)^2}\sqrt{GMa}
\end{equation}
where $\alpha$ is a scaling factor and the subscript \textit{inf} indicates that it is the specific orbital angular momentum injected at infinity into the system. \citeauthor{bate1997ballistic} and \citeauthor{bate1997gas} concluded that, in the terminology of $\eta$, $\eta>1$ when the initial specific angular momentum of the accreted material is smaller then the orbital specific angular momentum of the secondary.
Increasing the initial specific angular momentum of the accreted material to values above the specific orbital angular momentum of the secondary makes $\eta<1$.
Their argument is that small initial specific angular momentum material tends to fall directly towards the protobinary's center of mass and is thus accreted more easily by the primary.
Increasing the specific angular momentum of the infalling material, it forms a circumbinary disc around the system. In that case, material tends to fall towards the secondary protostar, as it is closer to the inner edge of the circumbinary disc. Figure \ref{fig:eta_q_estimates} however, reveals a large scatter for the models of \citet{bate1997gas}.

\citet{ochi2005} repeated the study of \citet{bate1997gas} with increased resolution of the simulations. \citeauthor{ochi2005} found that the primary protostar accretes the most material independent of \jinf. 
Table 1 in \citet{ochi2005} show simulations where the mass ratio evolution is towards both 1 and 0. 

The values of $\eta$ plotted in Fig.~\ref{fig:eta_q_estimates} shows a trend of $\eta$ increased scatter with the initial binary mass ratio. 
The models from \citet{young2015b} are fewer and show a constant value of $\eta$ for different mass ratios. 

From Fig. \ref{fig:eta_q_estimates}, deducing a systematic as to what determines the mass ratio evolution seems impossible. Although a direct comparison between the estimate of each study is not straight forward, it seems that there is no firm conclusion as to what $\eta$ should be. Basically, for the data presented in Fig. \ref{fig:eta_q_estimates}, $\eta$ is strongly dependent on the initial conditions of the gas, which are \textit{a priori} assumed. 
 
In summation, the value $\eta$ is largely unknown and our brief review of the hydrodynamic simulations suggests $-12<\eta<7$ which covers a larger interval than our estimated values discussed in 2.3 which are  between 1 $\leq \eta  \leq$ 5. A positive point however is that the larger interval obtained from hydrodynamical simulation at least include the smaller one deduced from simpler considerations.

\section{Modelling a single accreting pre-main sequence protostar}\label{sec:single_star}
Here we briefly describe our setup of MESA to simulate the pre-MS accretion phase and compare our results with relevant literature on accreting single pre-MS stars.

\subsection{The initial models}
We use MESA revision 8845 where we adopt the nuclear network \textit{pp$\_$extras.net} which includes reactions rates for Deuterium (D) burning. We adopt a standard solar metallicity $Z = 0.02$ with the standard solar composition of \citet{grevesse1998}. As modifications to this standard setting we set as initial mass fractions $X_{H_1}=0.70$, $X_{D} = 2.75\times 10^{-5}$ \citep[high primordial D/H ratio;][]{linsky2006}, $X_{He_3}=2.98\times 10^{-5}$, and $X_{He_4}=0.28$. For low temperature opacity we adopt the tables of \citet{freedman2008}. In order to explore a wide range in $q$, which is relevant in Sect. \ref{sec:grids}, we constructed a set of initial models using MESAs' \textit{create$\_$pre$\_$mains$\_$sequence$\_$model}-test-suite, where we have relaxed the initial mass from 1$M_{\odot}$ down to any mass in the range [0.1, 1]$M_{\odot}$ in steps of $0.01M_{\odot}$.

We start our simulations from a stellar structure with uniform composition. We assume that the composition of the accreted material is equal to the initial composition of the protostar and we ignore stellar rotation and mass loss from winds.

\subsection{Accretion rates}\label{sec:CH_law}
We will explore two types of mass-accretion rates onto the protobinary system which are then split and accreted onto either one of the two protostars. One is a constant accretion rate of $\dot{M} = 10^{-4}$\msun{}$\rm yr^{-1}$ and the second one is the so called CH accretion law. \citet{churchwell1997} proposed that the observed massive outflows found in high density HII regions were bipolar jets formed from infalling material of which a part was accreted onto the central object and another part was returned to the environment via outflows. Shortly thereafter this idea was independently confirmed by \citet{henning2000}. \citet{behrend2001} parameterized a relation between the mass outflows and the bolometric luminosty from the work of \citet{churchwell1997},
\begin{equation}\label{eq:CHlaw}
    \log \dot{M}_{out} = -5.28 + \log \frac{L}{L_{\odot}}\times \left(0.752-0.0278\log \frac{L}{L_{\odot}}\right) \, ,
\end{equation}
and then argued that, since the outflow was a fraction of the material falling towards the central object, another fraction $f_{acc}$ was accreted onto the central object. Hereby they get the simple relation that the accretion rate of the central object is
\begin{equation}\label{eq:CHdotM}
    \dot{M} = \frac{f_{acc}}{1-f_{acc}}\dot{M}_{out} \, ,
\end{equation}
where $f_{acc} = \frac{1}{11}$, which has been shown to reproduce the upper envelope of intermediate-mass single pre-MS stars \citep{haemmerle2016}, and the rates are measured in units of $\rm M_{\odot}yr^{-1}$.

The CH accretion law produces a variable accretion rate that increases with luminosity of the object. Its characteristic therefore is an increasing accretion rate, such that the more massive, hence luminous, an object becomes, the more mass it accumulates per unit time.

\subsection{Simulation of an accreting pre main sequence protostar}
\begin{figure}
    \centering
    \includegraphics[width=1.0\linewidth]{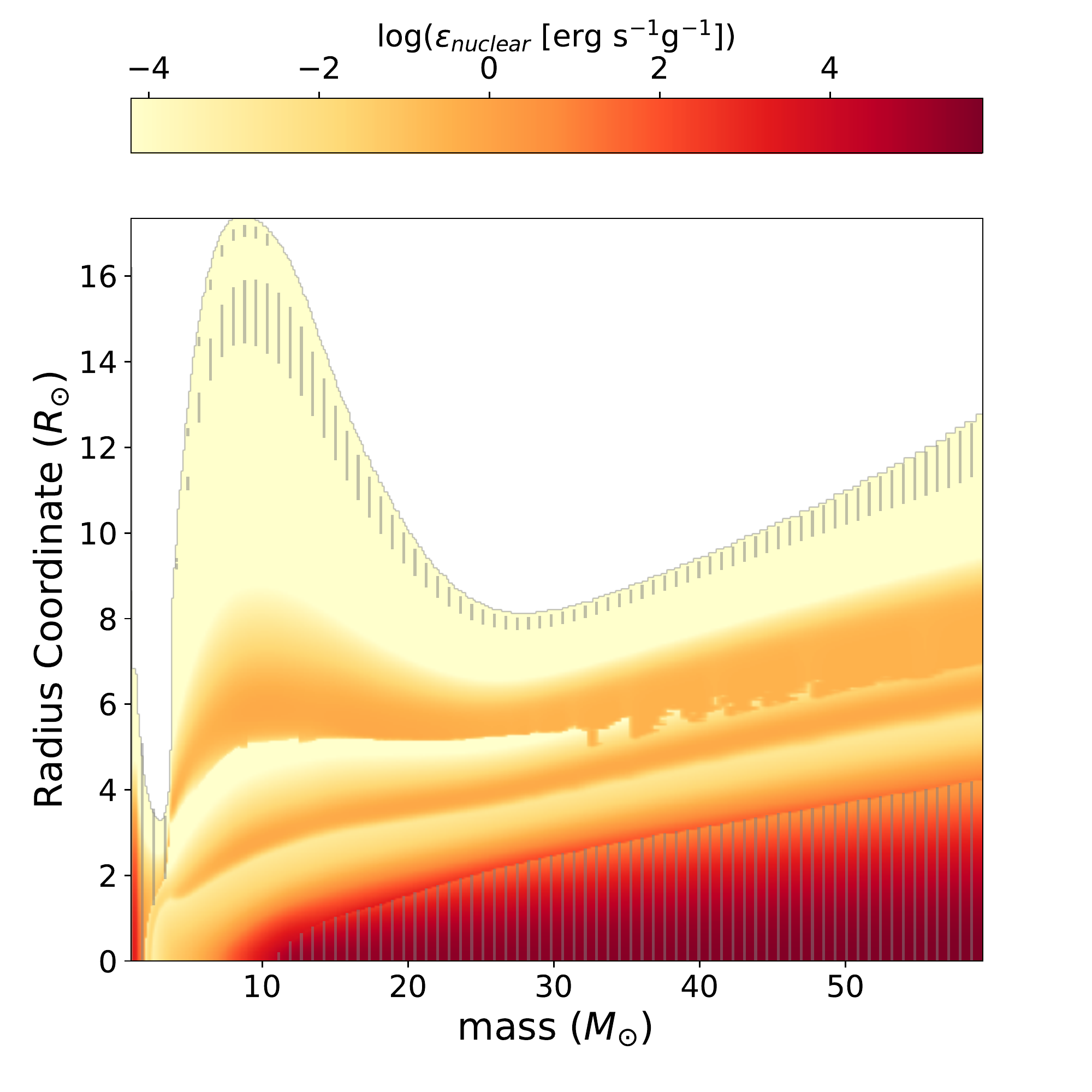}
    \caption{A protostar's Kippenhahn diagram showing its stellar radius as a function of mass. The protostar is undergoing Churchwell-Henning accretion growth at a metallicity Z=0.02. The color displays the specific power from nuclear reactions within the protostar. Gray zones are convective regions. The four different burning zones from surface to center are: D, $^{7}$Li, $^{3}$He , and H respectively.}
    \label{fig:eps_nuc}
\end{figure}

Figure \ref{fig:eps_nuc} shows a Kippenhahn diagram with the stellar radius coordinate as Y-axis and the total mass of the protostar undergoing Churchwell-Henning accretion at a metallicity Z=0.02 and deuterium mass fraction of $X_{D}=2.75\times10^{-5}$. The color displays the specific power from nuclear reactions within the protostar. We identify 4 burning zones during the pre-MS accretion phase. These are from the surface and inwards; D, $^{7}$Li, $^{3}$He , and H respectively. The grey zones are regions of convection. Accretion starts from a protostar of 1\msun{} and continues until it reaches 60\msun. The protostars evolution is in agreement with similar simulations in the literature, \citep[e.g.][]{haemmerle2016}. In the context of pre-MS binaries, the important element to note is the expansion due the core transforming from a convective to a radiative core and subsequently the shell D-burning leading to the peak seen in the radial expansion in Fig. \ref{fig:eps_nuc} around 10\msun{}. The expansion of the star happens because its core becomes radiative. A radiative core has a steeper densisty gradient relative  to the density gradient of the convective core. Hence, as the star becomes fully radiative, the lack of convection stops the transport of material into the core, and the star expands. The D-burning shell increases the effect of the expansion \citep{haemmerle2016}. We will refer to this phase of protostar evolution as the pre-MS swelling or simply the swelling phase.

\section{Grids of accretion growth pre-main sequence protobinaries}\label{sec:grids}
The problem of simulating the accretion phase of pre-MS protobinaries is challenging as it requires exploration of a high-dimensional parameter space. At least the following parameters, their time evolution, and associated processes, have an influence on the pre-MS and ultimately the post-ZAMS evolution of binary systems: $\eta$, $a_i$, $q_i$, $M_{1,ZAMS}$, $e_i$, initial spin of each protostar, and angular momentum content, tides, and internal energy of accreted material. Finally, it is possible that either of the two protostars overflow their Roche lobe leading to mass transfer and possibly mergers. Here we focus on non-rotating circular orbit protobinaries undergoing accretion via discs, hence we have 5 parameters of interest: $\eta$, accretion rate $\dot{M}$, mass of primary- and secondary protostar, and orbital separation. Of these, we explore in great detail $a_i$ and $q_i$.

In this section we present a comprehensive 2d-grid in $\log a_i$ and $q_i$ of accreting pre-MS binaries for different assumed values of $\eta=[-0.5, 0, 0.5, 1, 2]$. We vary the initial orbital separation in the range $\log a_i/R_{\odot} = [1; 6]$, using a step length $d\log a_i//R_{\odot} = 0.1$ when $\log a_i/R_{\odot} = [1, 2.5]$ while when $\log a_i/R_{\odot}>2.7$, the step length is $d\log a_i/R_{\odot}=0.5$. As initial mass ratios we have $q_i=[0.1;1]$ in steps of $dq_i = 0.01$. The initial primary mass is always $M_{1,i}=1M_{\odot}$. The grid contains a total of $\sim$128.000 models.

Of specific interest is the initial conditions of the protobinary that allow for the formation of a ZAMS binary with a specific primary mass $M_{1,ZAMS}$ for different accretion laws and values of $\eta$, while remaining detached. For each $\eta$ the splitting of mass was given from solving eq. \eqref{eq:qeta} under the constraint that $1=f_1 + f_2$. We use the CH accretion law, eqs. \eqref{eq:CHlaw} and \eqref{eq:CHdotM} and for sake of completeness produce for a $\eta=1$ a grid of models with a constant accretion rate of $10^{-4}$\msun{}$\rm yr^{-1}$. We define as stop criteria an upper mass of the primary star of $M_{1,ZAMS} = 60 M_{\odot}$ or if either of the two protostars overflow their Roche lobe. To generate the grid we have coupled our pre-MS accreting protostar model setup as described in Sect. \ref{sec:single_star} with a modified version of the MESA binary module to allow accretion onto two stars simultaneously \footnote{detailed setup of the runs, MESA inlist files, and the modified version of the MESA binary module will be available from $\rm http://cococubed.asu.edu/mesa_market/inlists.html$}.
In the next section, we will often refer to the grid with $\eta=1$ since the mass ratio remains constant, hence understanding this grid is simpler. However, this does not mean that n=1 is our preferred value; rather this is just the most pedagogical presentation of the grids.

\subsection{Minimum initial orbital separation}
\begin{figure}
    \centering
    \includegraphics[width=1.0\linewidth]{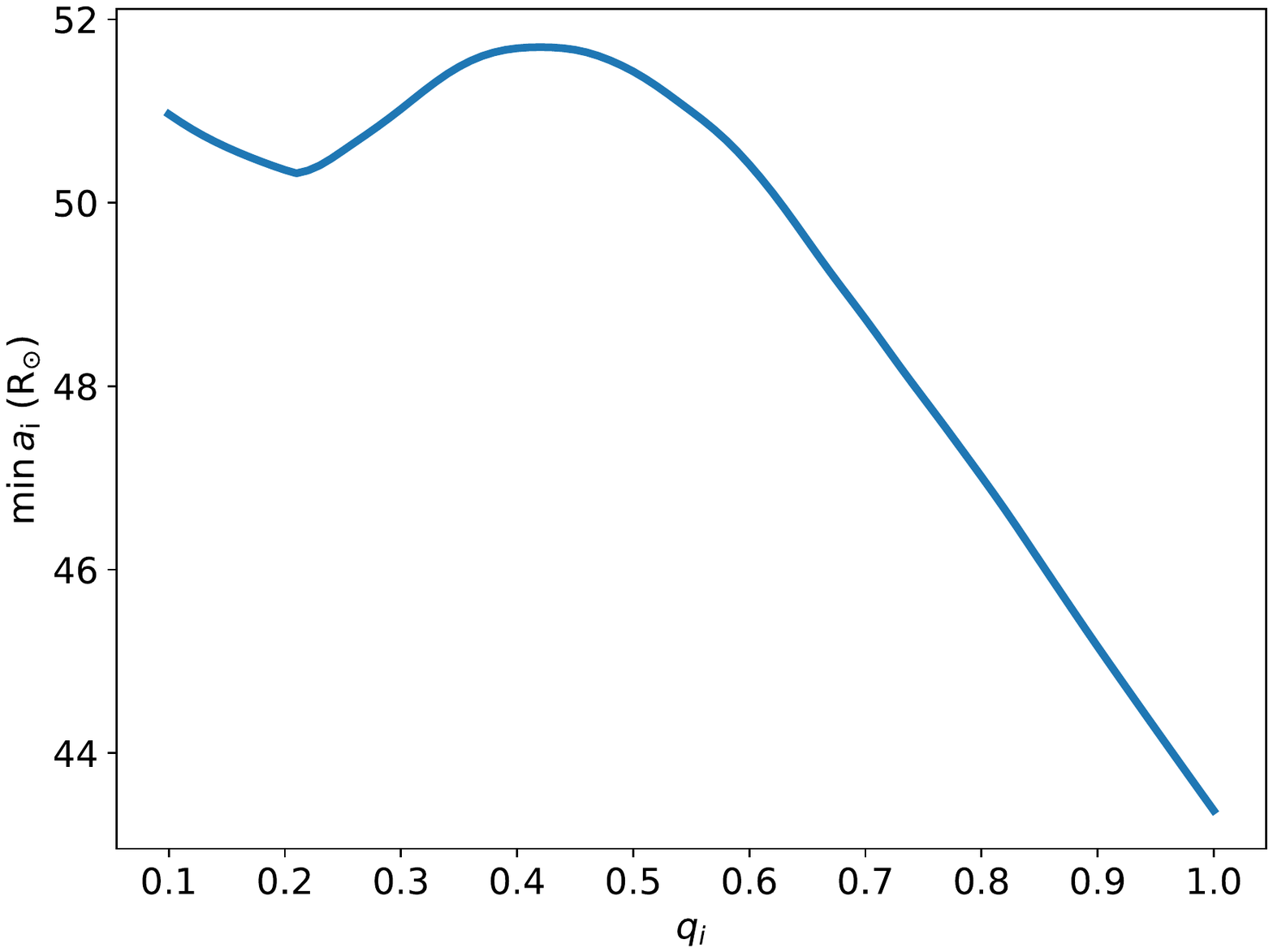}
    \caption{The minimum initial orbital separation required to allow the protostars to fit inside the Roche sphere as a function of initial mass ratio $q_i$ when the primary protostar is $M_{1,i}$=1\msun.}
    \label{fig:min_ai_vs_qi}
\end{figure}

The MESA pre-MS protobinary accretion sequences are setup in a way such that if the initial protostar in a binary overfills its Roche lobe the computation stops. This also means that there is a minimum initial orbital separation $\min a_{i}$ for our protobinaries as a function of initial mass ratio $q_i$. In Fig. \ref{fig:min_ai_vs_qi} we show this minimum initial orbital separation. Notice how, the minimum initial orbital separation increases with decreasing $q_i$.

\subsection{Examples of pre main sequence accreting protobinaries}
We show four indicative examples of pre-MS accreting protobinaries and qualitatively discuss important aspects of their evolution from the onset of accretion and until the sequence stops, either because one of the two stars overflow their Roche lobe or the primary mass reaches a target mass of 60 \msun. Given our purpose to explore the boundaries of detached protobinary accretion, we do not follow here the post-accretion phase when the two stars contract on the ZAMS. Actually this phase will change neither the mass ratio nor the orbital period of the system.

Figures \ref{fig:M1_RLO_D-shell} to \ref{fig:M2_RLO_low_mass} each contains a Hertzsprung-Russel (HR) diagram on the left panel and the radius, Roche radius, and orbital distance with time in the right panel of the two protostars. The HR diagram displays the stellar luminosity of each proto star, not to be mistaken for the luminosity from accretion. Actually, this accretion phase, in general cannot be observed since the two stars are still hidden inside the cloud that provides the material that falls onto the protobinary system. On the other hand, when the accretion stops, the two stars becomes visible at the last point of these lines (that are called for this reason birthlines). The last point is the one when the maximum mass is reached (see the masses indicated along the birthlines).

\subsubsection{The primary protostar overflows the Roche Lobe}
\begin{figure*}
    \centering
    \includegraphics[width=1.0\linewidth]{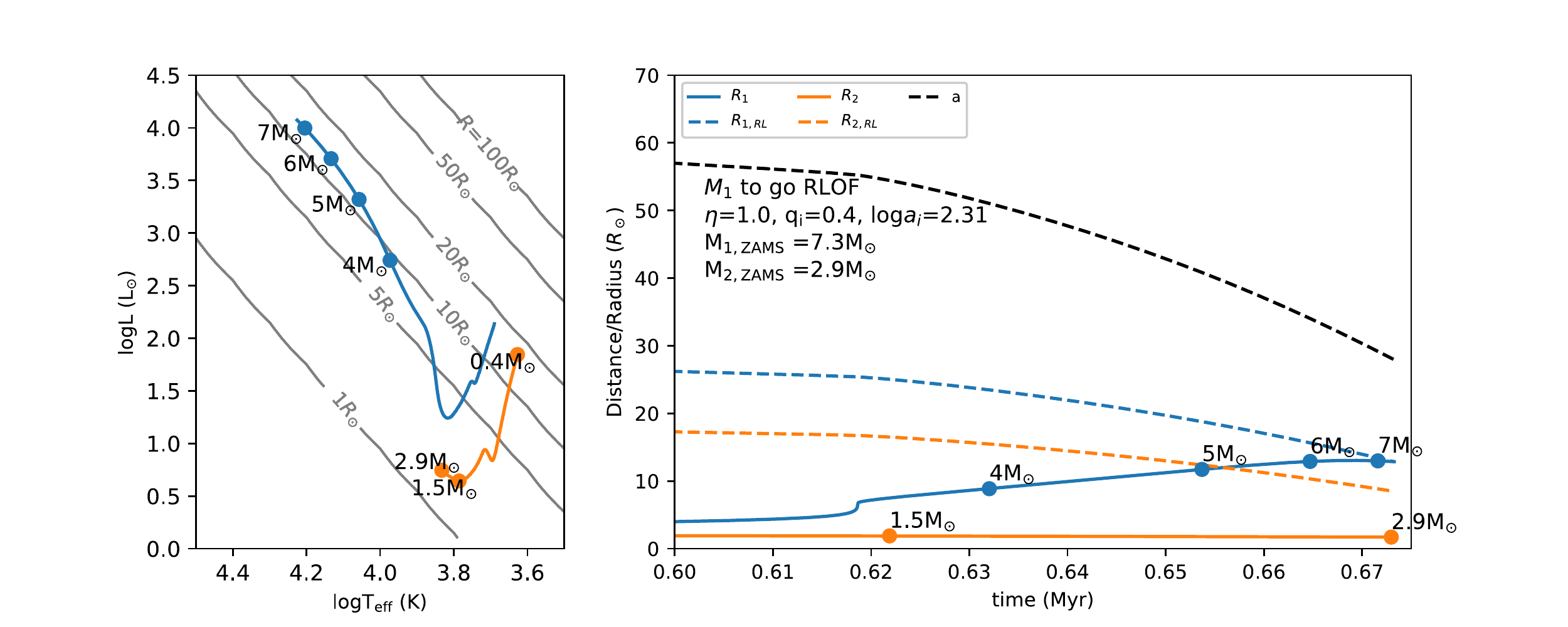}
    \caption{Example of a pre-MS protobinary accretion sequence, where the primary star fills its Roche lobe during the swelling phase. Blue color denotes the primary star and orange color denotes the secondary star. The left panel shows the HR diagram for the two binary components' evolution. Constant radius lines are shown in gray for radii of 0.1, 1, 10, 50, and 500\rsun. The right panel shows the evolution of the radius (solid) of the two stars, their Roche radius (dashed), and the orbital separation (black dashed) as a function of time. The masses of the two stars at selected times are shown as labelled dots along the solid lines.}
    \label{fig:M1_RLO_D-shell}
\end{figure*}
The first example is a protobinary undergoing accretion growth with a rate given by the CH-accretion law, and with initial orbital separation $a_i=10^{2.31}\simeq204$\rsun{}, $M_{1,i}=1$\msun{}, initial mass ratio $q_i=0.4$, and $\eta=1$.
Figure \ref{fig:M1_RLO_D-shell} shows, in the left panel, the HR diagram of the two protostars, and the right panel shows, the evolution of the radius of each protostar, their Roche radii, and orbital distance as a function of time. Blue lines refer to the primary star and orange lines to the secondary star, solid lines indicate parameters of the protostars, while dashed lines are the Roche radii of each star, and the black dashed line is the binary's orbital separation. 
The computation stopped when the primary star filled its Roche lobe. At this point in time, the binary system has increased its total mass from 1.4\msun{} to 10.2\msun{} with the binary mass ratio being unchanged.
In this accretion sequence the binary system increased its mass by a factor $\sim7$. Increasing the initial orbital separation would allow the binary system to grow to higher masses.
From the right panel, it is seen, that as the binary system grew in mass, the orbit and the two Roche radii have shrunken. At the end of the computation, the orbital separation has decreased to $\sim$30\rsun. Around 0.62 Myr, the primary star enter the swelling phase, increasing its radius until the computation stops.
In the HR diagram, we note that the primary star's has increased its luminosity by 4 orders of magnitude during the accretion phase.
The behaviour of the secondary star in the HR diagram, shows it has undergone an initial contraction down to a minimum luminosity and minimum radius. The initial contraction gave an increase in effective temperature $T_{\rm eff}$.

\begin{figure*}
    \centering
    \includegraphics[width=1.0\linewidth]{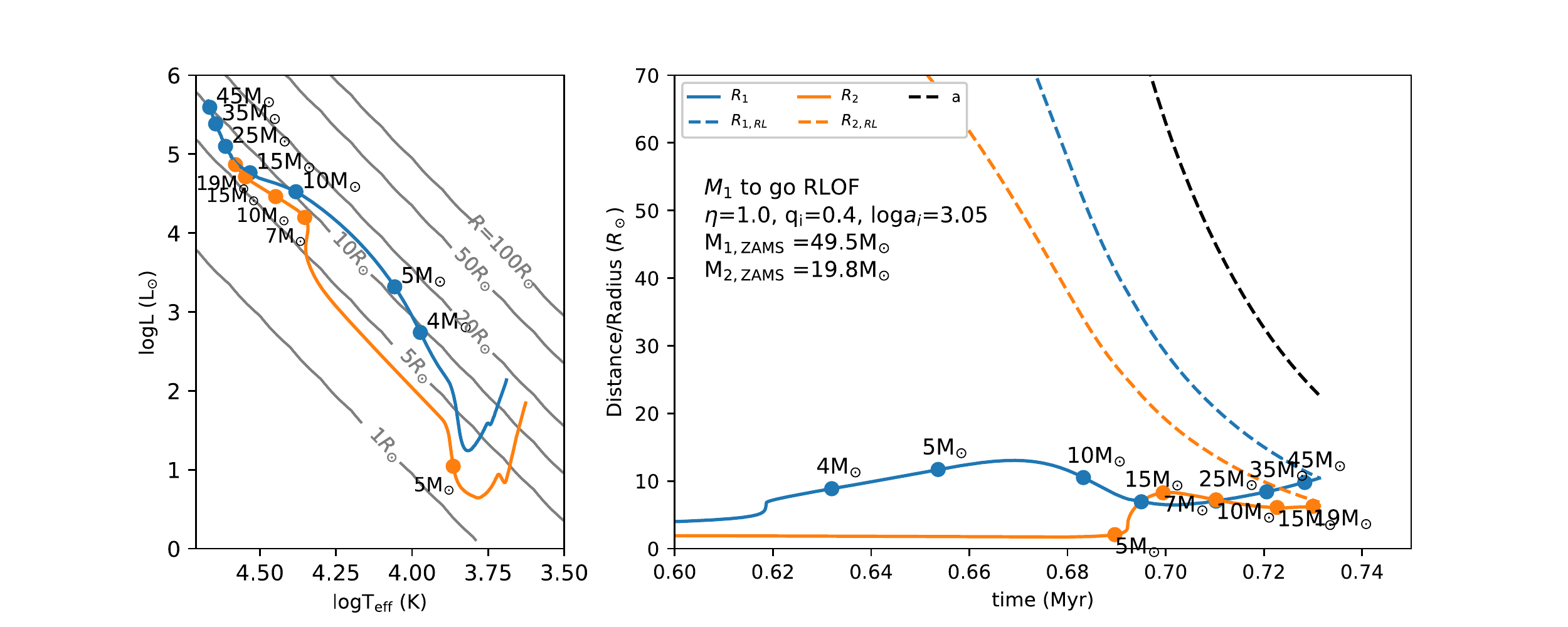}
    \caption{Example of a pre-MS protobinary undergoing accretion growth where the primary star overflows its Roche lobe upon beginning its second expansion phase and after contracting from the swelling phase. Layout of the figures is the same as in Fig. \ref{fig:M1_RLO_D-shell}.}
    \label{fig:M1_RLO_high_mass}
\end{figure*}
The second example was similar to that of Fig. \ref{fig:M1_RLO_D-shell}, but with the initial orbital separation increased to $a_i=10^{3.05}\simeq1122$\rsun{} and is displayed in Fig. \ref{fig:M1_RLO_high_mass}. The figure reads the same way as Fig. \ref{fig:M1_RLO_D-shell}. 
Again, the computation stopped when the primary star filled its Roche lobe. At this point in time, the total mass of the binary system has increased from 1.4\msun{} to 69.3\msun{}. Note the initial orbital separation is only some $60\%$ larger over the first example above. But the system's total mass has increased by a factor of $50$. The final orbital separation is again around 30\rsun{} like in Fig. \ref{fig:M1_RLO_D-shell}. 
The evolution of both stars are similar to that described in Fig. \ref{fig:M1_RLO_D-shell}, but the slightly larger initial orbital separation, allow the system grow to a higher mass. We see that both stars have grown beyond the ZAMS and begun core Hydrogen burning.
Looking at the HR diagram first, on the left panel, the two protostars display different evolutionary paths. The primary star, blue, is very similar to a single star undergoing accretion growth as simulated by \citet{haemmerle2016}. Like the primary star, the secondary star in orange, increases its luminosity and $T_{\rm eff}$, but the small accretion rate, means this star takes much longer to initiate its swelling phase, and in the mean time its radius remains nearly constant during the accretion.
The stellar radius of the secondary star is, like its birthline on the HR diagram much different compared to the primary star. The primary star spends a much longer period undergoing the swelling phase relative to the secondary star. The secondary star on the other hand went through the swelling phase, expanding less than the primary star.
The radial evolution of the two stars is such that, had they been slightly closer initially, the secondary star would have overflown its Roche lobe during its swelling phase. 

For a binary star to reach a mass $\gtrsim$10\msun{}, i.e. avoid overflowing its Roche lobe during the accretion phase, it must  have 
an initial separation larger than some limit. If the initial separation is just beyond this limit its final mass reachable from accretion increases significantly, see also Figs. \ref{fig:minLogP_0.5} and \ref{fig:minLogP_2.0}.

\subsubsection{The secondary protostar fills its Roche lobe}
\begin{figure*}
    \centering
    \includegraphics[width=1.0\linewidth]{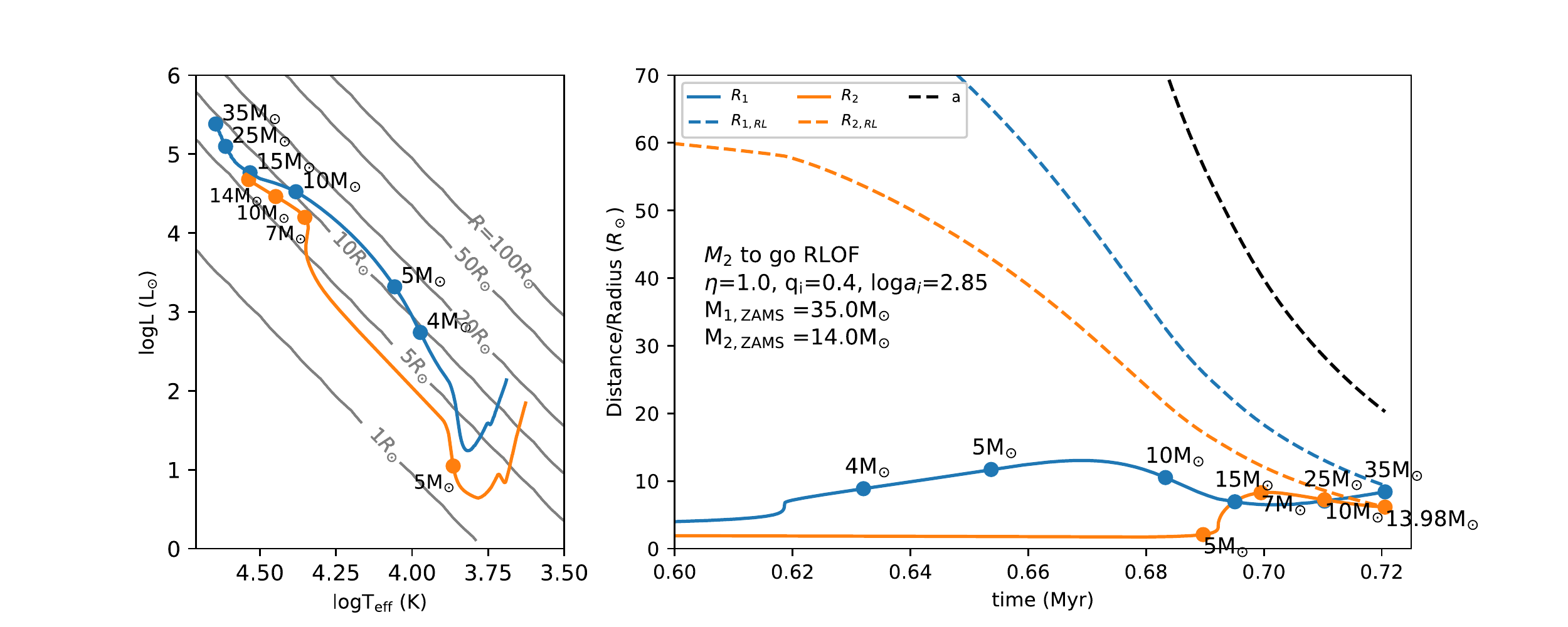}
    \caption{Example of a pre-MS protobinary in accretion growth where the secondary star fills its Roche lobe when it expands in the swelling phase. Layout of the figure is the same as in Fig. \ref{fig:M1_RLO_D-shell}.}
    \label{fig:M2_RLO_D-shell}
\end{figure*}

\begin{figure*}
    \centering
    \includegraphics[width=1.0\linewidth]{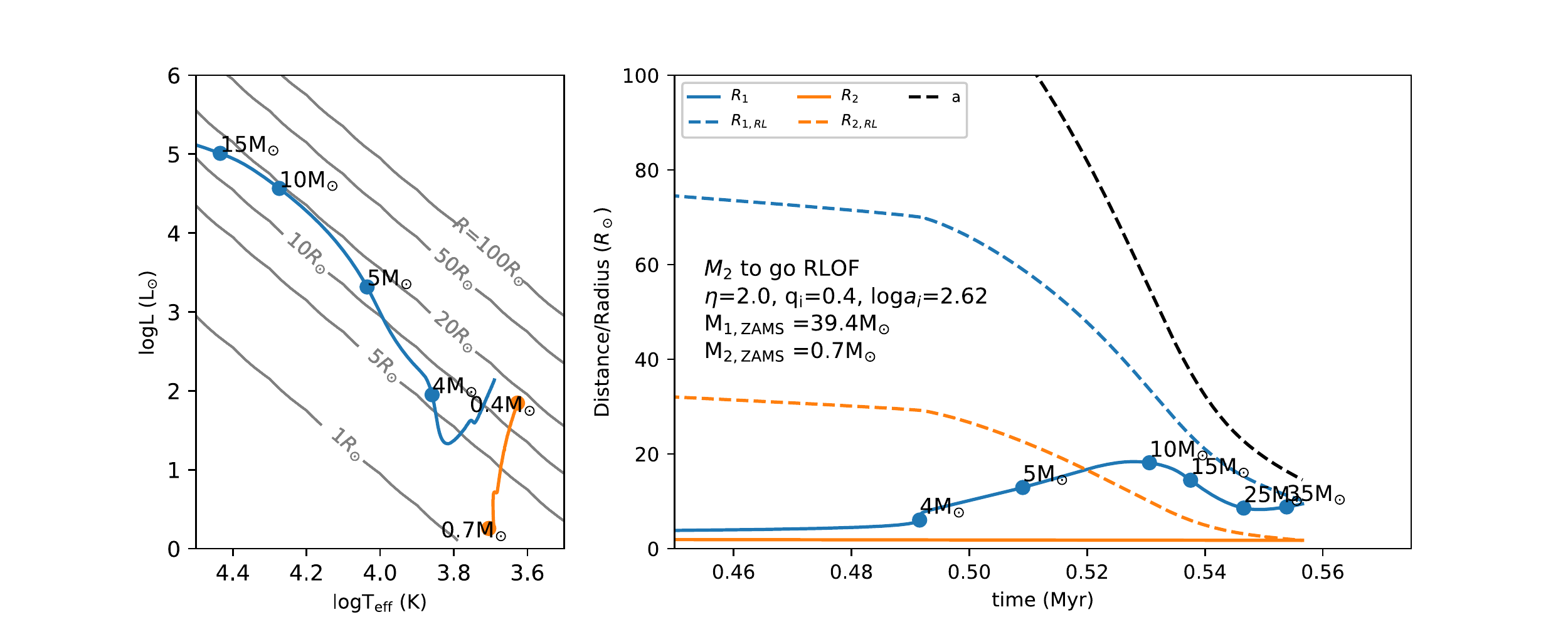}
    \caption{Example of a pre-MS protobinary in accretion growth where the secondary star fills its Roche lobe. Layout of the figure is the same as in Fig. \ref{fig:M1_RLO_D-shell}.}
    \label{fig:M2_RLO_low_mass}
\end{figure*}

We identify two ways in which the secondary star fills its Roche lobe and each are illustrated in Figs. \ref{fig:M2_RLO_D-shell} and \ref{fig:M2_RLO_low_mass}.
In the third example, shown in Fig. \ref{fig:M2_RLO_D-shell}, the binary has an initial orbital separation of $a_i=10^{2.85}\simeq708$\rsun{}, identical to the example shown in Fig. \ref{fig:M1_RLO_D-shell}. In this third example, the combination of initial mass ratio and initial orbital separation is such that the accretion sequence stops as the secondary protostar fills its Roche lobe during the swelling phase. In fact there is a whole range of initial mass ratios and orbital periods in which the secondary star will fill its Roche lobe. We elaborate on this in Figs. \ref{fig:grid_min_aiM1zams_eta1p0} and  \ref{fig:grid_min_aiM1zams_eta2p0}.

The fourth and final example is shown in figure \ref{fig:M2_RLO_low_mass} and shows the evolution of two protostars undergoing accretion growth with a rate from the CH-accretion law. Its initial orbital separation is $a_i=10^{2.62}\simeq417$\rsun{}, it has primary protostar of $M_{1,i}=1$\msun{}, a binary mass ratio $q_i=0.4$, and $\eta=2$. The evolution of the primary star is already well described in the previous three examples.
The secondary star's behaviour is however different, because its accretion rate is small, and the star never grows above 0.7\msun{}. The small accretion rate results in a track that is similar to constant mass evolution.
The convective star follows the Hayashi limit downwards, with a nearly constant Teff.
The mass grows slowly, and thus the star contracts slowly, on a timescale of low-mass stars, see left panel of Fig. \ref{fig:M2_RLO_low_mass}. The small accretion thus slows the contraction of the star. The small mass growth of the secondary star gives the binary a very small mass ratio. As the binary system grows in mass, the secondary star's Roche lobe decreases, and eventually it closes around the secondary star that fills the Roche lobe. Even though mass transfer from a smaller to a larger mass increases the orbital separation, the continued accretion onto the primary star which is significant at this late stage means that the primary star will consume the secondary star or if the accretion stopped, leave the system with a substellar companion. But the system, might be driven by other physics as well, which we will explore in a different work.


\subsubsection{Example of 2D grid for $\eta = 1$ and $\eta=2$}\label{sec:ex_2dgrid}
\begin{figure*}
    \centering
    \includegraphics[width=1.0\linewidth]{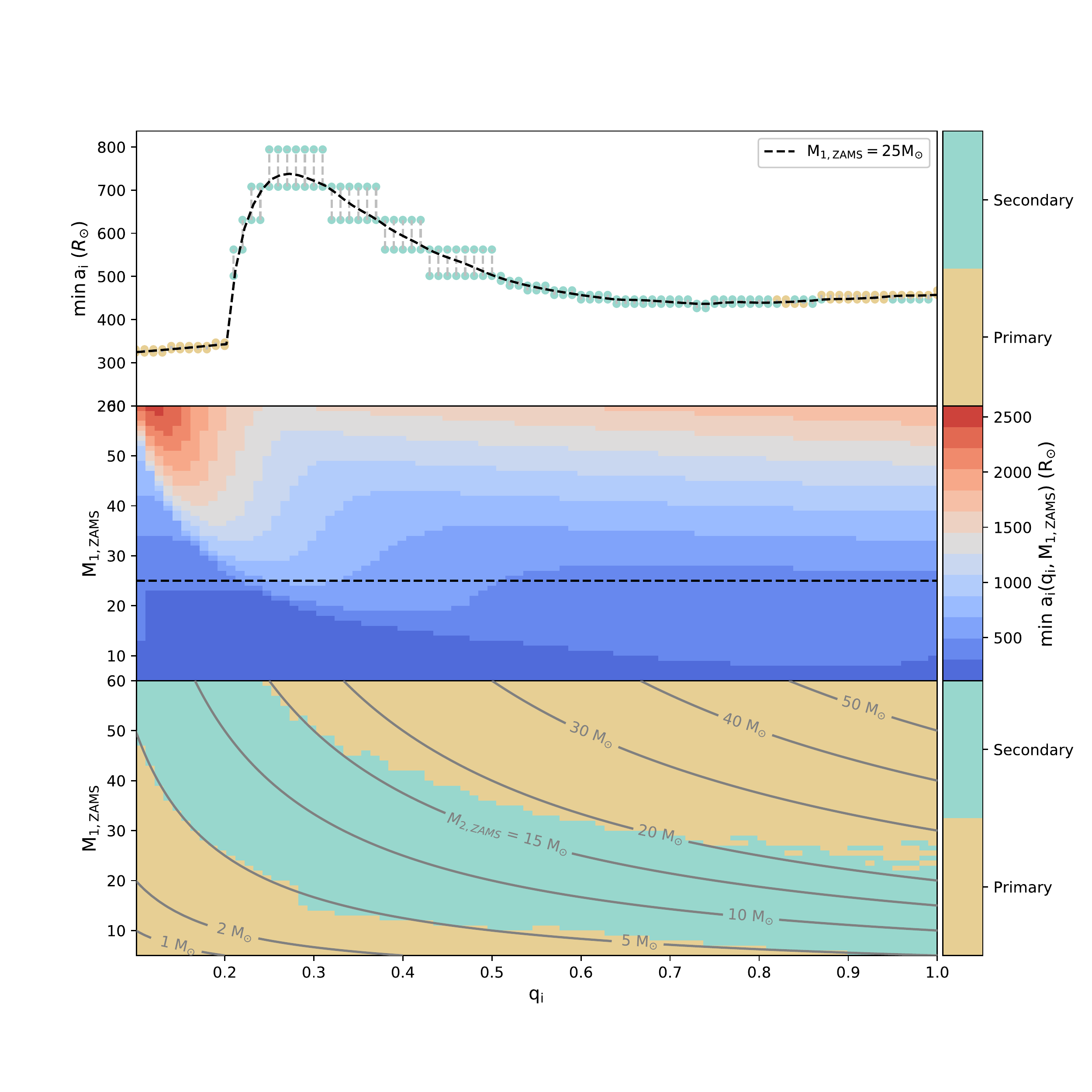}
    \caption{2D grid of pre-MS accretion growth protobinaries with $\eta=1.0$ and the CH-accretion law. \textbf{Top panel}: The dashed line is the minimum initial orbital separation, $\min a_i$, as a function of the initial mass ratio $q_i$ required to form a primary star at ZAMS of $M_{1,ZAMS}=25M_{\odot}$. Filled circles are MESA pre-MS accretion sequences used for interpolation from which the dashed line is generated (see text). A pale green color indicates that the MESA sequence stopped because the secondary protostar entered Roche lobe overflow. A pale brown color signals that the primary protostar went Roche lobe overflow. \textbf{Middle panel}: The initial minimum orbital separation projected into the $M_{1,ZAMS}$ vs $q_i$ plane. 
     On the first axis is shown the initial mass ratio $q_i$ and on the second axis is the primary mass at ZAMS $M_{1,ZAMS}$ in units of $M_{\odot}$. The color indicates the corresponding minimum initial orbital separation required to reach a specific mass of the primary star at ZAMS. Dashed line is at constant $M_{1,ZAMS}=25$\msun. \textbf{Bottom panel}: Distribution of primary/secondary protostar to overflow its Roche lobe in the $M_{1,ZAMS}$ vs $q_i$ plane if accretion continues beyond $M_{1,ZAMS}$, with similar color description as in the top panel. The contour indicates levels of constant secondary star mass, $M_{2,ZAMS}$, at ZAMS.}
    \label{fig:grid_min_aiM1zams_eta1p0}
\end{figure*}

Figure \ref{fig:grid_min_aiM1zams_eta1p0} shows our 2D-grid in $\log a$ and $q_i$ of accreting pre-MS protobinaries for $\eta=1$. This grid is the easier of the six $\eta$-values to understand because $q$ remains constant during the accretion phase.

The dashed line in the top panel in Fig. \ref{fig:grid_min_aiM1zams_eta1p0} shows the minimum orbital separation required to form a primary star of $M_{\rm 1,ZAMS}$=25\msun{} while being a detached binary, as a function of initial mass ratio $q_i$. 
As a numerical example, to form a 25 M$_\odot$, starting from a mass ratio equal to 0.28 (and ending also with that mass ratio since with $\eta$=1, $q$ remains constant), implies a minimum orbital distance of around 720 R$_\odot$.
The two green colors linked by a dashed vertical line associated to this same value of the initial ratio (at 690 and 790 R$_\odot$) have the following meaning: the system starting with an initial distance of 690 (790) R$_\odot$ allows forming a less (more) massive primary star than 25 M$_\odot$ because the secondary reaches the Roche Limit.

A striking feature appearing in the upper panel of Fig. \ref{fig:grid_min_aiM1zams_eta1p0} is the fact that the minimum initial separation needed to form a given mass in a detached accreting scenario does not vary monotonously with the initial mass ratio. A naive first guess would have expected that the higher the mass ratio, the larger the initial separation. Actually we see that the curve shows a sharp increase around $q_i=0.2$, passes through a maximum and then decreases, until it increases again very slowly when $q_i$ is higher than 0.7.

The sharp increase around $q_i$=0.2 is a consequence of the secondary star's swelling phase.
If $M_{\rm 1,ZAMS}=25$\msun{}, Fig. \ref{fig:eps_nuc} shows  that the secondary protostar has a larger radius than the primary protostar when its mass is between 5-25\msun. 
Of course the primary had also to evolve along this swelling phase, but this did not allow it to
reach the Roche limit because the secondary had at that time a smaller mass and the two stars had a larger separation. What allows now the Roche limit to be reached is the fact that two stars
are nearer from each other since they both have grown in mass.
Roche lobe filling during the secondary star's swelling phase is not a fine tuned configuration but is easily obtained for a broad range of initial mass ratios and initial orbital periods. 

The middle panel of Fig. \ref{fig:grid_min_aiM1zams_eta1p0} shows the minimum initial orbital separation $\min a_i$ in the ($q_i$, $M_{\rm 1, ZAMS}$)-plane. The dashed line in the middle panel is at $M_{1,ZAMS}=25$\msun{} that was shown in the top panel.
For each protobinary with initial mass ratio $q_i$, the color indicates the lower limit of initial orbital separation needed, in order for that protobinary to reach some primary ZAMS mass $M_{\rm 1,ZAMS}$ through accretion growth, while remaining detached. As a numerical example, to form a primary ZAMS mass $M_{1,ZAMS}$ = 60\msun\, for an initial mass ratio $q_i=0.13$ the initial orbital separation is $a_i\simeq2550R_{\odot}$. So the middle panel outlines the lower limits of initial orbital separations and initial mass ratios required to build a specific star of some mass at ZAMS. 
This panel can also be used, given initial conditions, to predict whether the system might reach, through the accreting detached scenario, some final mass.
For instance, we can forecast that the ZAMS primary mass of IRAS 04191+1523 \citep{lee2017} (its current orbital separation of $860$ au $\simeq10^{5.26}$\rsun{} and mass ratio is $\simeq 0.85$) can reach through accretion masses of 30-40 \msun{} while remaining detached.

Overall, the greater the primary mass at ZAMS, the greater the initial orbital separation is needed, to avoid a star filling its Roche lobe before accretion stops. However, note the peak in $\min a_i$ which is at $q_i$ = 0.13 and $M_{1,ZAMS}$ = 60 \msun{} and not at a higher $q_i$.
The bottom panel can explain the overall trend of which protobinary component will overflow its Roche lobe first. The pale brown color says, that, if for a model sequence in that region, accretion continues beyond a specific primary mass at ZAMS, given an initial mass ratio, the primary protostar will fill its Roche lobe. The pale green color indicates that for continued accretion beyond some ZAMS mass, given an initial mass ratio, the secondary protostar will fill its Roche lobe first. The grey lines are lines of constant secondary star mass at ZAMS. We should note the green band spans values of $M_{2,ZAMS}$ in the range 5-20\msun. This is exactly the mass range in which the secondary stars is undergoing expansion and contraction, going through the swelling phase, hence has its largest radial extend. The green band thus, is a product of the secondary protostar's Roche lobe being only a fraction of the primary protostar's Roche lobe and the swelling phase of the secondary star.
Thus, having the secondary star fill its Roche lobe during pre-MS is easily achieved from the present stellar models. This is different from the conclusions of \citet{krumholz2007} who considered constant high accretion rates forming stars up 100\msun. A high constant accretion rate means that the swelling phase happens at a larger mass, compared to our models, which also means, that the interplay between Roche lobe filling, orbital separation changes and the sweilling phase, produces a different output. Thus, whether Roche lobe filling of the secondary star happens in nature depends on the distribution of protobinaries at the onset of accretion and the exact accretion history.

\begin{figure*}
    \centering
    \includegraphics[width=1.0\linewidth]{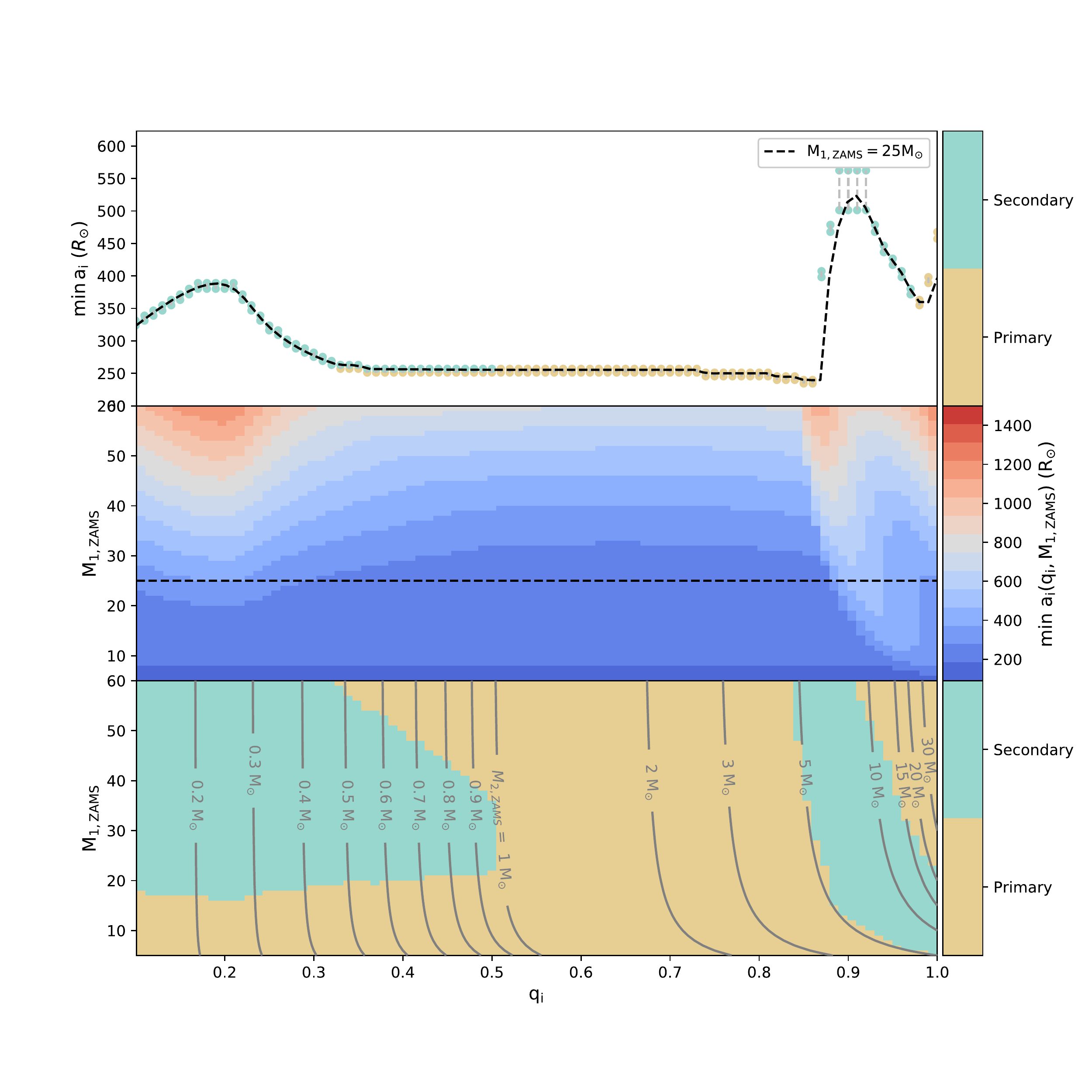}
    \caption{Same as Fig. \ref{fig:grid_min_aiM1zams_eta1p0} for $\eta = 2$. We identify the right pale green region as the same pale green region seen in Fig. \ref{fig:grid_min_aiM1zams_eta1p0}. The bottom panel shows a new green region compared to bottom panel of Fig. \ref{fig:grid_min_aiM1zams_eta1p0} which are the systems described in Fig. \ref{fig:M2_RLO_low_mass} where the secondary protostar ends fillings its Roche lobe.}
    \label{fig:grid_min_aiM1zams_eta2p0}
\end{figure*}

Figure \ref{fig:grid_min_aiM1zams_eta2p0} is identical to Fig. \ref{fig:grid_min_aiM1zams_eta1p0} but with $\eta=2$. With this value of $\eta$, the mass ratios evolve and tends towards lower values. In general the orbital separation required for accretion up to 60\msun{} primary star is smaller. The reason is that, for a given $q_i$, increasing $\eta$ decreases the amount of mass accreted by the binary to reach a specific primary star mass (for a given increase of the primary mass, less mass goes into the secondary). 
The band of models where the secondary star has masses between 5 and 20 M$_\odot$ in Fig. \ref{fig:grid_min_aiM1zams_eta1p0} bottom panel, is where the secondary star fills it Roche lobe. In Fig. \ref{fig:grid_min_aiM1zams_eta2p0}, this same band is also seen, but shifted to higher initial mass ratios.
A new feature in Fig. \ref{fig:grid_min_aiM1zams_eta2p0} over Fig. \ref{fig:grid_min_aiM1zams_eta1p0} is the region of secondary stars filling their Roche lobe at very small mass ratios for ZAMS primary masses above 17\msun. If accretion stopped just at the moment before Roche lobe overflow, these systems would have extremely small mass ratios and are qualitatively similar to the evolutionary sequence illustrated in Fig. \ref{fig:M2_RLO_low_mass}.

\subsection{Minimum orbital period at ZAMS with $M_{\rm 1,ZAMS}$ and $q_{\rm ZAMS}$}
\begin{figure}
    \centering
    \includegraphics[width=1.0\linewidth]{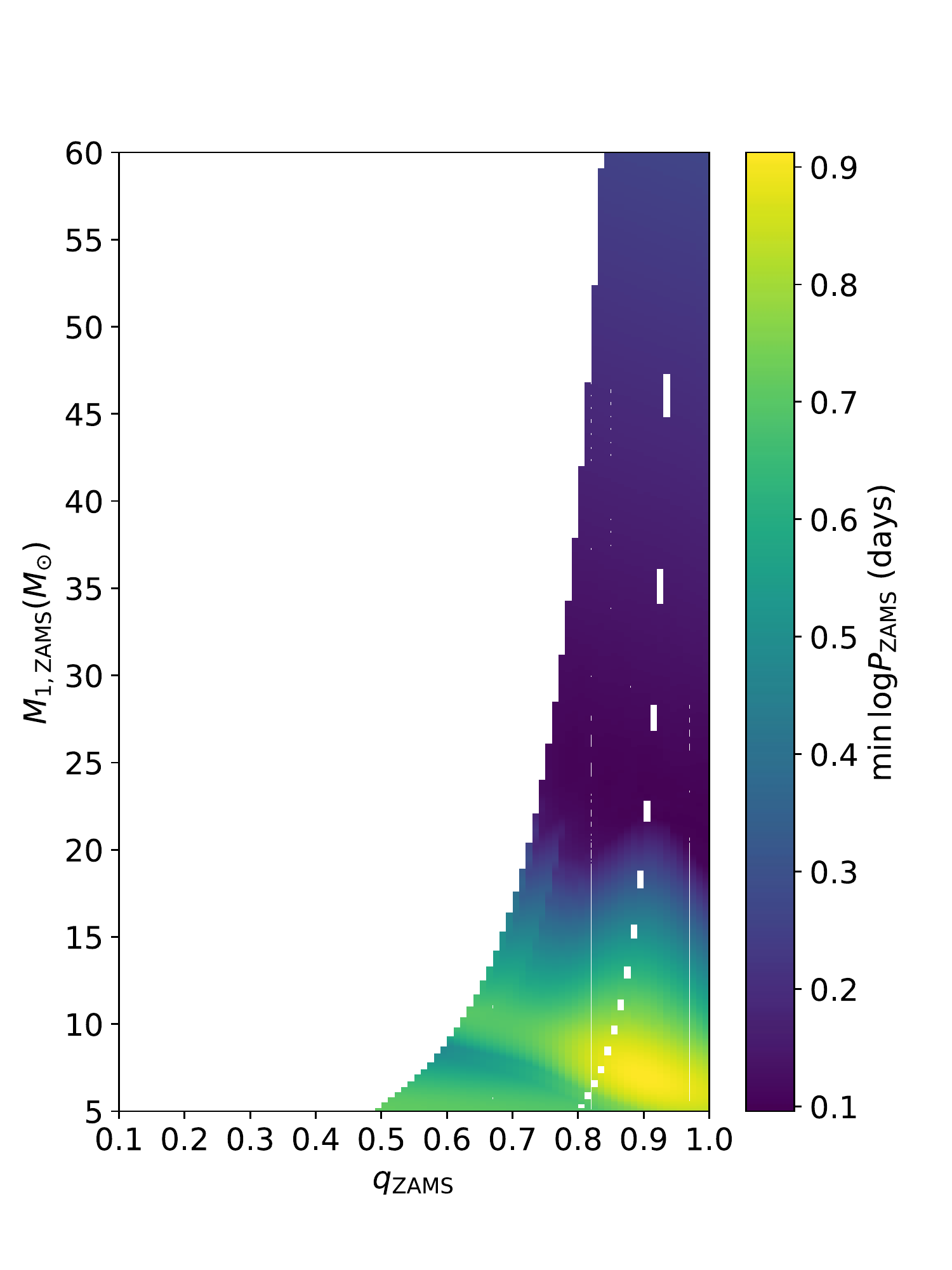}
    \caption{Minimum orbital period at ZAMS in the $M_{\rm 1,ZAMS}$-$q_{\rm ZAMS}$-plane when $\eta=0.5$, for binary systems forming through accretion and remaining detached. The white areas correspond to parts of the parameter space which is not accessible by detached evolution.}
    \label{fig:minLogP_0.5}
\end{figure}

\begin{figure}
    \centering
    \includegraphics[width=1.0\linewidth]{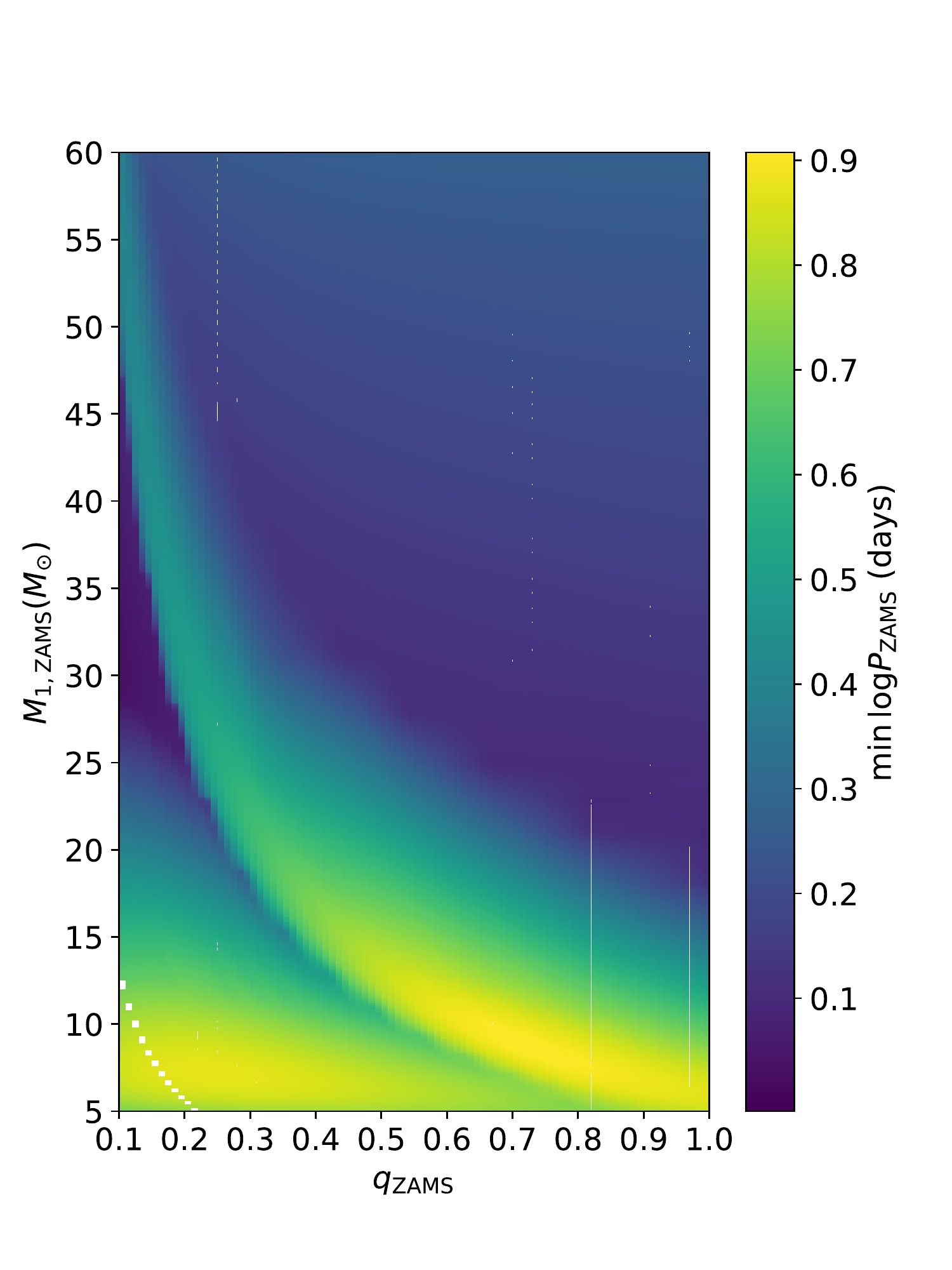}
    \caption{Minimum orbital period at ZAMS in the $M_{\rm 1,ZAMS}$-$q_{\rm ZAMS}$-plane when $\eta=2$, for binary systems forming through accretion and remaining detached.}
    \label{fig:minLogP_2.0}
\end{figure}

Another important piece of information that we extracted from our grid of models is the minimum orbital period of binary systems that can form from a phase of detached accretion during the pre-MS.
In Fig. \ref{fig:minLogP_0.5} we show the minimum orbital period, $\log P_{ZAMS}$, as a function of $M_{1,ZAMS}$ and $q_{ZAMS}$ when $\eta=0.5$ and Fig. \ref{fig:minLogP_2.0} shows the same function when $\eta$=2.0. 

When $\eta$=0.5, the mass ratio increases as a function of time and thus the systems that can be obtained are pushed towards mass ratios on the ZAMS that are above 0.5 for the range of the primary masses considered in the plot. In systems with a ZAMS primary mass equal to 60 \msun{}, the mass ratios on the ZAMS cover a smaller range than in systems containing a smaller primary mass on ZAMS. This is consistent with the analytical approach given previously (see $q_f$-limit given by eq. \eqref{eq:limqf} and Fig.\ref{fig:eta_qi2qf}). When $\eta$=2.0, the mass ratio decreases as a function of time and the mass ratio on the ZAMS can cover a larger range of values than in the case $\eta$=0.5.

In all accretion sequences the separation decreases as a function of accretion growth. This means that the orbital period also always decreases as a function of accretion growth. So the minimum orbital period on the ZAMS, corresponds to a minimum initial orbital period allowed for detached binary evolution to that point on the ZAMS. The minimum initial orbital period is determined by the situation occurring along the accretion process where the two stars just graze its Roche limit. 
As seen in Sect. \ref{sec:ex_2dgrid}, it can be either the primary or secondary star that fills the Roche lobe first. In situations where only the primary can fills its Roche lobe (this is for cases where the mass ratios at ZAMS are high), the behaviour of the minimum orbital period, more or less follows the evolution of the radius as a function of mass seen in Fig. \ref{fig:eps_nuc}. When the secondary star is approaching the Roche limit first (i.e. for mass ratios at the ZAMS below about 0.8 for the mass range considered in Figs. \ref{fig:minLogP_0.5} and \ref{fig:minLogP_2.0}), the minimum orbital period on ZAMS, does not vary monotonically with primary mass, given a fixed ZAMS mass ratio. This non-monotonicity results from the fact that in order for the secondary to avoid reaching the Roche limit during the swelling phase, the minimum initial orbital period (and hence the minimum ZAMS orbital period) has to be increased. This is what breaks in this figures the smooth evolution of the colors.

If we compare the minimum orbital periods obtained for a 5\msun{} primary (whatever the value of $q_{\rm ZAMS}$) with the values of the minimum orbital periods for 60\msun{} primary star, we see that smaller ZAMS orbital periods are obtained for the most massive primaries. Although, the system constaining the 60 M$_\odot$ has to start with a larger initial orbital period than the system containing the 5\msun, the separation decreased more in the case of the 60\msun{} than in the case of the 5\msun{} and thus a smaller ZAMS orbital period at the end for are obtained for the system containing the 60\msun{} primary star.
The minimum ZAMS orbital period for stars above 40\msun{} and high mass ratios could even proceed to evolve via the chemical homogeneous evolution channel. A potential interesting candidate for such a system is the massive overcontact binary VFTS352 which has an orbital period of just 1.1241452(4) day \citep{almeida2015} and equal mass components of $~28$\msun. This system, being just on the limit of that inferred possible, for such a system in Fig. \ref{fig:minLogP_0.5} and \ref{fig:minLogP_2.0}, could likely have formed from scenario of close massive binary formation presented here. Since VFTS352 is in the 30 Dor star forming region within LMC, its metallicity is smaller compared to our adopted solar metallicity as stars at lower metallicity has typically smaller radii. This might allow for the formation of slightly more tight close binaries.
It might also be, as is mentioned in \citet{almeida2015}, that the system is currently undergoing chemical homogeneous evolution and is thus, due to internal mixing, more compact, than a non-rotation models we have made here. However, \citet{almeida2015} find the system is relative unevolved hence it is close to its end-of-accretion settings.

\section{Results}\label{sec:result}

\subsection{Comparisons with observed distributions of $P_{\rm rot}$ and $q_{\rm ZAMS}$, case of 40 M$_\odot$}\label{sec:comp}

Here we explore our models' ability to explain the observed distribution of binaries. The distribution of binary systems is composed of 4 dimensions, i) primary mass $M_1$, ii) mass ratio $q$, iii) orbital period, $\log P$, typical in log scale, and  iv) eccentricity $e$. Recently \citet{moe2017} deduced the joint probability distribution $p(M_1, \log P, q, e)$ which is different from the product of multiplying individual probability distributions. The \citeauthor{moe2017}-distribution includes observations that spans the range of primary masses 0.8<$M_1/M_{\odot}$<40, mass ratios between 0.1 and 1, and $\log (P/\rm days)$ = [0.2:8]. In the primary star mass range this has been extrapolated to primary star masses up to 100 \msun{}. The distribution also includes eccentricities which we do not consider in this paper, since at present our theoretical framework also neglects eccentricity. Thus, whenever we use the empirical joint distribution by \citeauthor{moe2017}, we first marginalize it over the eccentricity.

In order to compare our models with the \citeauthor{moe2017}-distribution function, we sample, for specific primary masses between 6 and 50 \msun{}, a binary population of $2\times10^6$ systems that follows the mass ratio and orbital period distribution given in \citet{moe2017} at that primary mass. We define these binary systems as our observed ZAMS binary systems. An analogy would be to propose a model for single star formation and validate this model against its ability to reproduce the inital stellar mass function (IMF). 

\begin{figure*}
    \centering
    \includegraphics[width=0.5\linewidth]{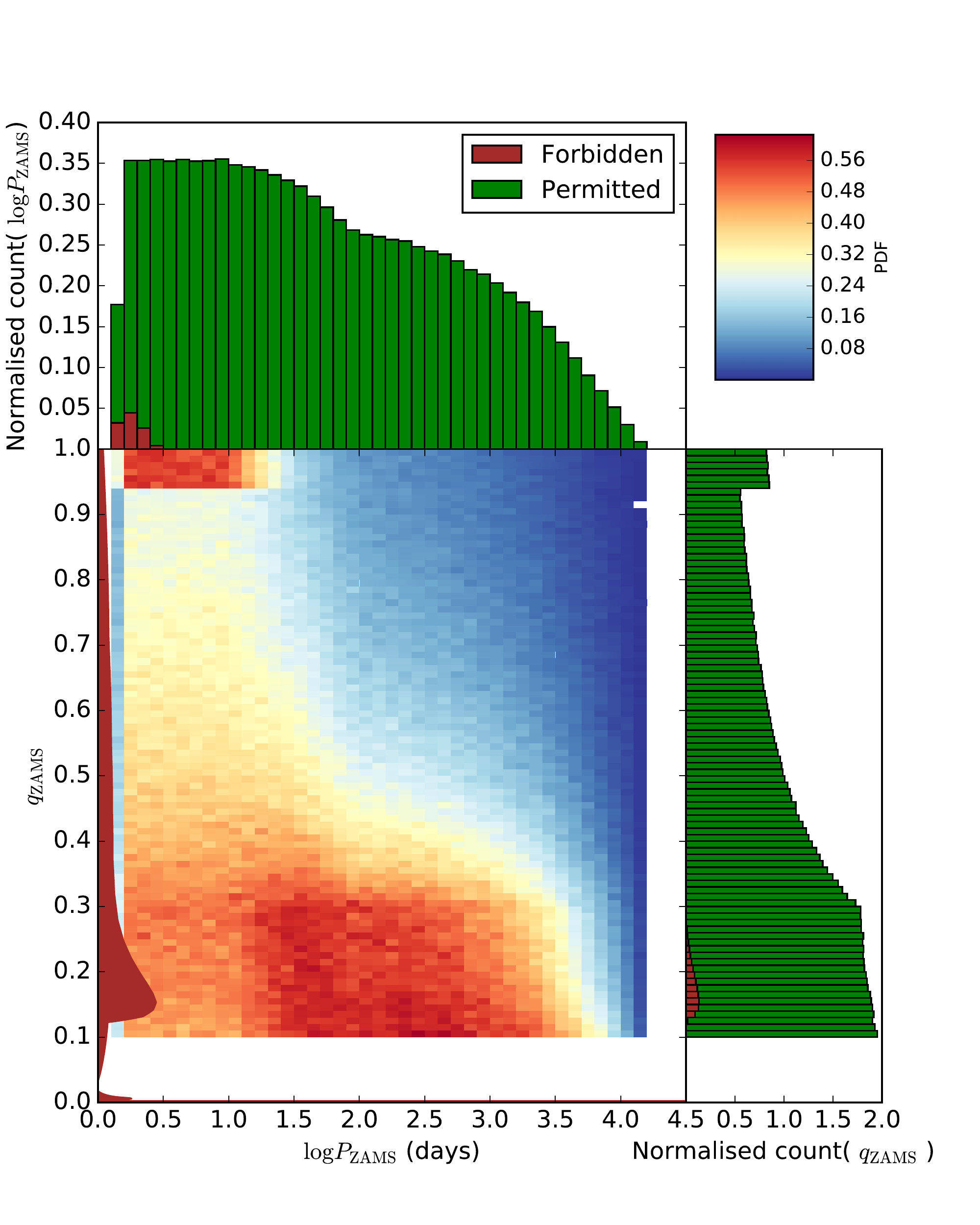}\includegraphics[width=0.5\linewidth]{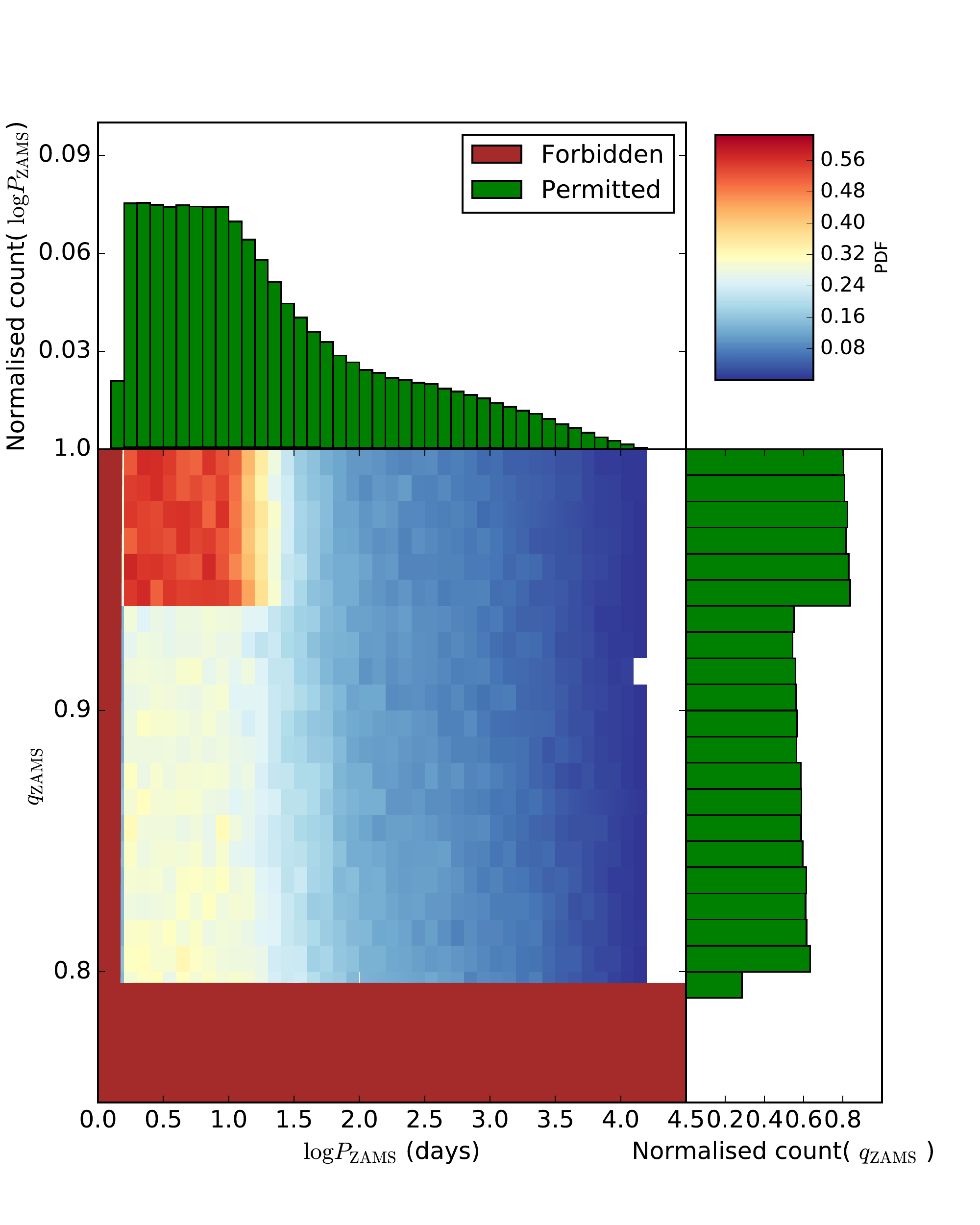}    
    \caption{Distribution of binary systems in mass ratio $q$ and orbital period $\log P$ for primary stars of $M_{1, ZAMS}=40$\msun{} for all eccentricities from \citet{moe2017}. \textbf{2D color plot:} normalised 2D distribution of mass ratios and orbital periods. \textbf{Top panel:} The normalised distribution in $\log P$. \textbf{Right panel:} The normalised distribution in $q$. The brown regions indicate systems that cannot be reached by detached evolution during the accretion phase. {\it Left Panel:} A value $\eta$=2.0 is used. {\it Right Panel:} A value $\eta$=0.5 is used.}
    \label{fig:q_logP_Mprim40eta2}
\end{figure*}

For each primary mass at ZAMS, we represent its distribution of binary systems in the mass ratio and orbital period as a normalised 2D histogram. An example is shown as a three-panel-plot in Fig. \ref{fig:q_logP_Mprim40eta2} where the color plot is the 2 dimensional distribution in orbital period $\log P/$days on the first axis and mass ratio $q$ on the second axis. The color bar indicates the density of binary systems in each bin. The two histograms on the edges show the marginalised $\log P$-distribution (top) and $q$-distribution (right) respectively. Figure \ref{fig:q_logP_Mprim40eta2} shows the $q$-$\log P$-distribution for a primary star of 40\msun. We note that at this primary mass, most systems are found with mass ratios below $q<0.4$ and with orbital periods between 0.2 < $\log (P/\rm days)$ < 3.5. Finally we note a small excess in mass ratio above $q\geq0.95$, also known as the twin excess population.

The question that we now ask is what fraction of these observed ZAMS binaries has an orbital period and mass ratio which allow them to have been formed through an accretion process, as described in this paper, for different values of the parameter $\eta$? To answer this question we overlay the limit of $\min \log P$ as a function of mass ratio $q_{ZAMS}$ for a specific primary star mass $M_{\rm 1, ZAMS}$ in the three panel plot shown in Fig. \ref{fig:q_logP_Mprim40eta2}. In essence, this is drawing a vertical line in the 2d histogram and counting the number of binary stars to the right and left, respectively of the vertical line. Binary systems to right of the line, can have formed via accretion, as a detached binary system and we call them \textit{permitted}. Systems to the left of the line cannot have formed through accretion while remaining detached, and we call these the \textit{forbidden} systems.


The three-panel-plots in Fig. \ref{fig:q_logP_Mprim40eta2} show two examples where we have masked out the forbidden region in the 2d histogram with a brown color, assuming $\eta=2$ and $\eta=0.5$, respectively. Shown as the 2D color plot is the mass ratio versus orbital period distribution for a primary star mass of 40\msun{} and we compare our models with $\eta=2$ and $\eta=0.5$.

In the left panel of Fig. \ref{fig:q_logP_Mprim40eta2}, where $\eta=2$, there is only a very small number of systems that cannot be reproduced from the accretion growth formation channel. The fraction concealed beneath the brown region is about 1\% of the full sample. We see these are short period systems with mass ratios around $q\sim0.2$. Comparing to the bottom panel of Fig. \ref{fig:grid_min_aiM1zams_eta2p0} we see this bump in the forbidden region corresponds to the mass range where the secondary star overflows its Roche lobe, due to the swelling of the secondary star.

In the right panel of Fig. \ref{fig:q_logP_Mprim40eta2} where $\eta=0.5$, the brown area covers most of the observed sample.  Judging from the right panels two side-panels of Fig. \ref{fig:q_logP_Mprim40eta2} showing the distribution of $q_{ZAMS}$ and $\log P_{ZAMS}$, it is not only most of the range in which we observe binary systems at this primary mass, it is also most of the systems that cannot be reproduced when $\eta=0.5$.

In fact, this model only recovers systems of high mass ratio. Naturally, we wonder whether this results is affected by the fact that we did not consider initial mass ratios below 0.1. It does not seem to be the case. Even using a very low initial mass ratio, \eqref{eq:limqf} and Fig.~\ref{fig:eta_qi2qf} tells that only systems with a mass ratio larger than about 0.5-0.6 can be formed.


\subsection{Conditions at the beginning of the accretion phase, case of 40 M$_\odot$}\label{sec:inicond}

\begin{figure*}
    \centering
    \includegraphics[width=0.5\linewidth]{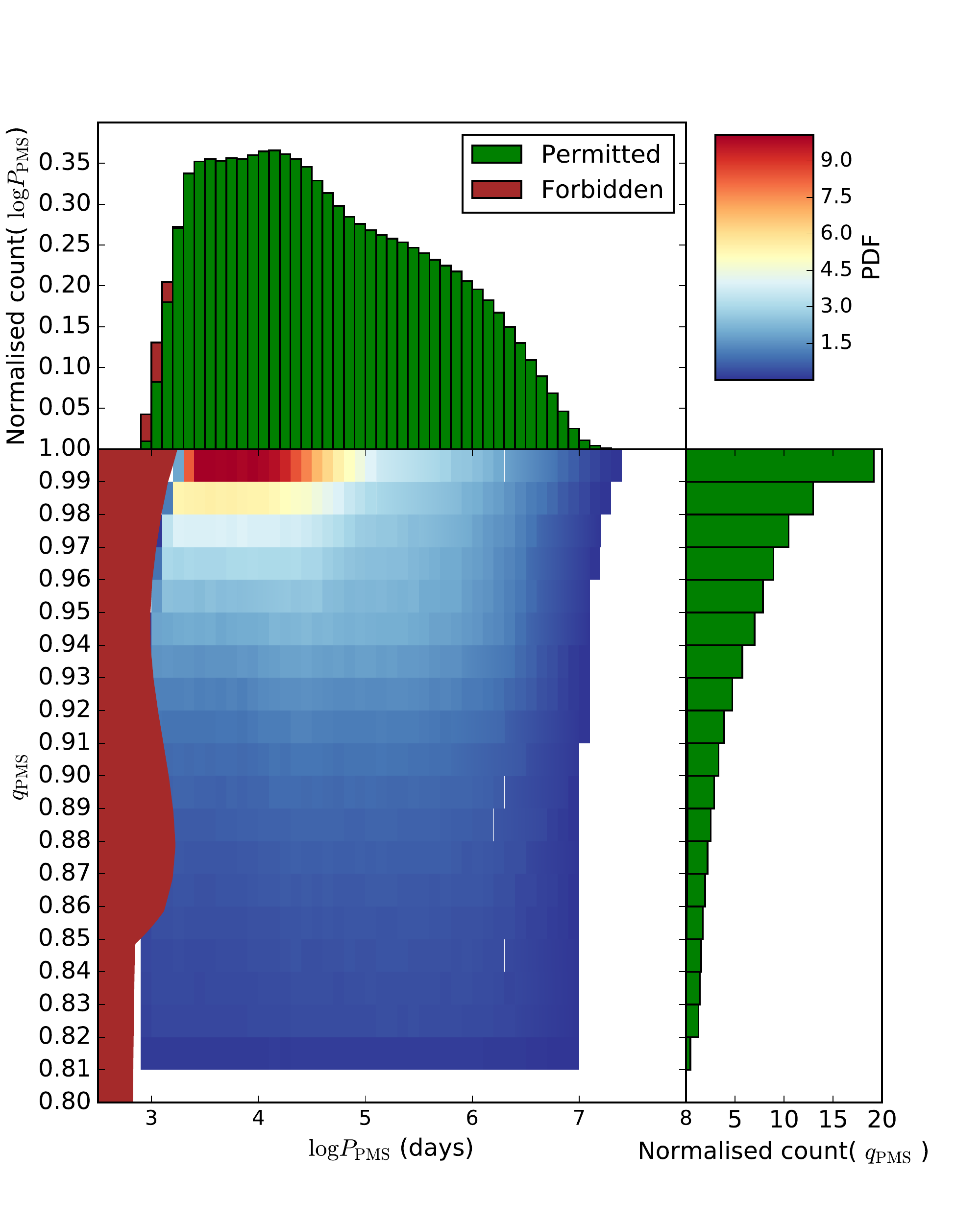}\includegraphics[width=0.5\linewidth]{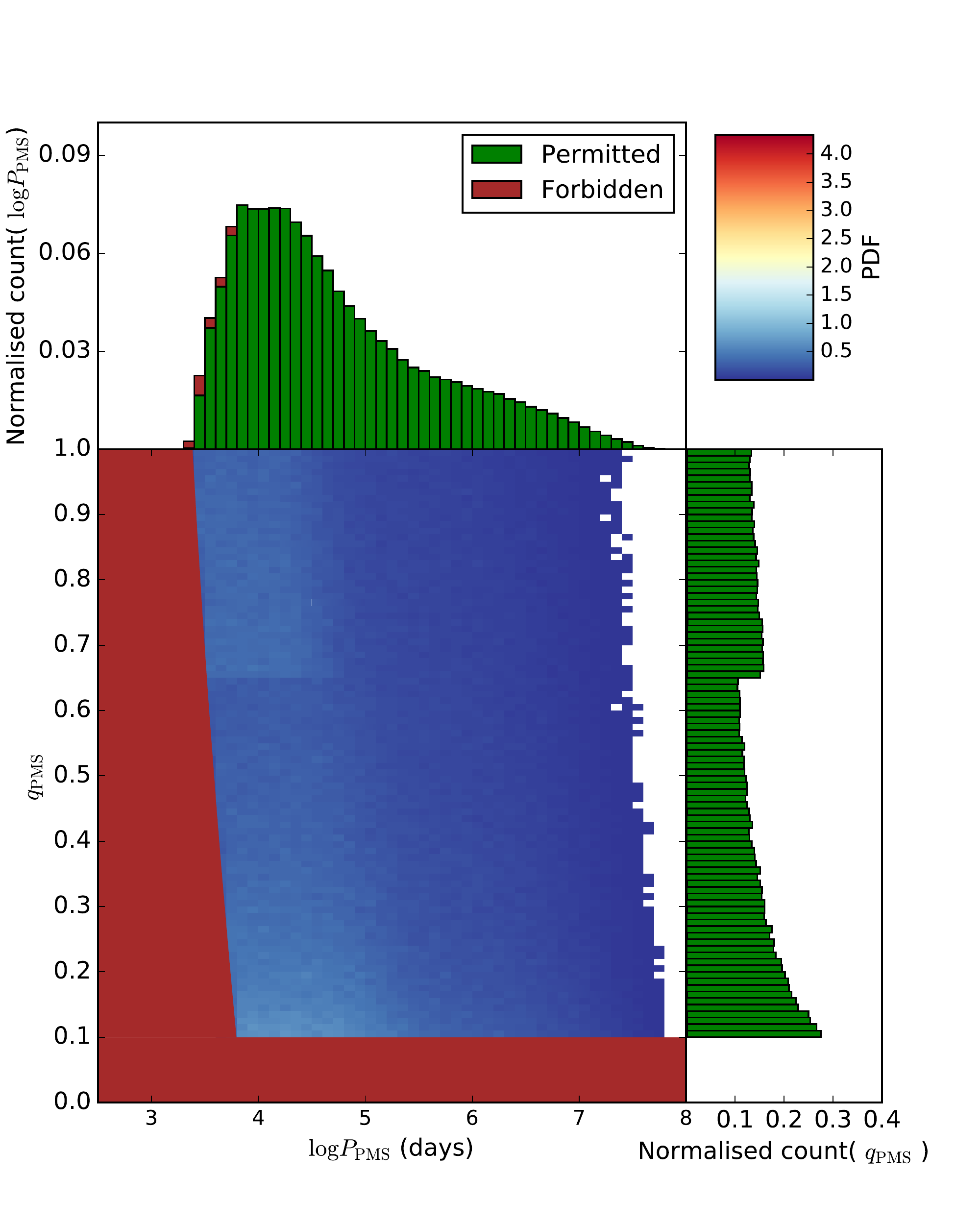}
    \caption{Distributions of mass ratio $q$ and orbital period $\log P$ for protobinary stars at the onset of accretion that can lead to the distribution of binaries at ZAMS with a primary star of $M_{1, ZAMS}=40$\msun{}, shown in Fig. \ref{fig:q_logP_Mprim40eta2}, assuming in the left figure that $\eta=2$ and in the right figure that $\eta=0.5$. \textbf{2D color plot:} normalised 2D distribution of mass ratios and orbital periods. The brown area is the region inside which a protobinary is unable to grow a primary star of 40\msun{} from detached accretion grow when $\eta=2$ and covers only $\sim1\%$ of the full distribution. Note the systems have high mass ratios above $q_i=0.8$ when accretion starts. \textbf{Top panel:} The normalised distribution in $\log P_{PMS}$. \textbf{Right panel:} The normalised distribution in $q_{PMS}$. Forbidden regions are labelled brown. Permitted regions are shown with green colors.}
    \label{fig:pms_q_logP_Mprim40eta2}
\end{figure*}

Using the MESA binary accretion sequence grids and the analytic relations derived in sect. \ref{sec:orbit_evo} and \ref{sec:mass_ratio_evo} we can infer  the required properties of the protobinary population so that the currently observed ZAMS binary properties are reproduced, under our proposed accretion-growth binary formation channel.
Examples of this initial protobinary distribution for $M_{ZAMS,1}=40$ and $\eta=2$ or $\eta=0.5$ are shown in the left and right panel of Fig. \ref{fig:pms_q_logP_Mprim40eta2} respectively.
As an example, we take a ZAMS binary system with a primary star mass $M_{1,ZAMS}=40$\msun{}, $q_{ZAMS}=0.5$ and $\log P_{ZAMS}=2.0$. To calculate its initial mass ratio $q_i$ and $\log P_i$ we do as follow: We assume that $\eta=2.0$ and that the protobinary has a primary protostar of initial mass 1\msun{}. Then, the initial mass ratio is directly given from eq. \eqref{eq:qf2qi} and comes out to be $q_i= 0.9756$. The orbital period is also straight forward to compute from eq. \eqref{eq:pfpi_disk} and comes out to be $\log P_i = 2.96$.
If $\eta=2.0$, the protobinary distribution at the onset of accretion contains a very large population of high mass ratio systems. This is expected since the mass ratio decreases when more and more mass is accreted. In that case,  what leads to the difference between the final masses of the two stars is an initially small mass difference between the two proto-stars. 

At least two mechanims might be able to produce near-equal mass proto-binaries at the onset of the accretion phase: The Jeans mass of the fragments that collapse in the molecular cloud is a product of the cloud's conditions at that point in its evolution. Thus two fragments collapsing nearly at the same time and become bound will have nearly the same mass. A second mechanism could be three body interactions in a dense star forming region, which would also tend to produce equal mass binaries at the expense of the smallest proto-star among the three.

Comparing the initial distribution of the orbital periods to the ZAMS one, we note that the initial distribution is shifted to larger periods by about 3 orders of magnitude. This is due to the accretion process which decreases the binary orbital separation. If we want to compare the overall shape of the two distributions, we need to normalise the periods. We can for instance normalise each period by the maximum initial period considered. Then we have that the bins at low normalised orbital periods on the ZAMS are filled at the expense of systems showing larger normalised initial orbital periods. This is expected since
many systems will be stacked in the lower bins after accretion.
Except for this feature, the general shape of orbital period distributions are nearly the same between the beginning of our computation and the ZAMS, indicating in that case, that the observed orbital period distribution is somewhat inherited from processes
occurring before accretion growth phase.

When $\eta$=0.5, the initial mass ratios cover the whole spectrum. This is also expected since in this case, the mass ratio is increasing when the accreted mass increases. The initial distribution of the orbital periods is also shifted to larger values with respect to the distribution obtained on the ZAMS. Comparing Fig.~18 and 16, we can do the same remarks as for the case computed with $\eta$=2.0. 
But this distribution is very far from the observed one due to the many systems that are forbidden in this case.

\subsection{The ZAMS orbital period distributions}

\begin{figure*}
    \centering
    \includegraphics[width=1.0\linewidth]{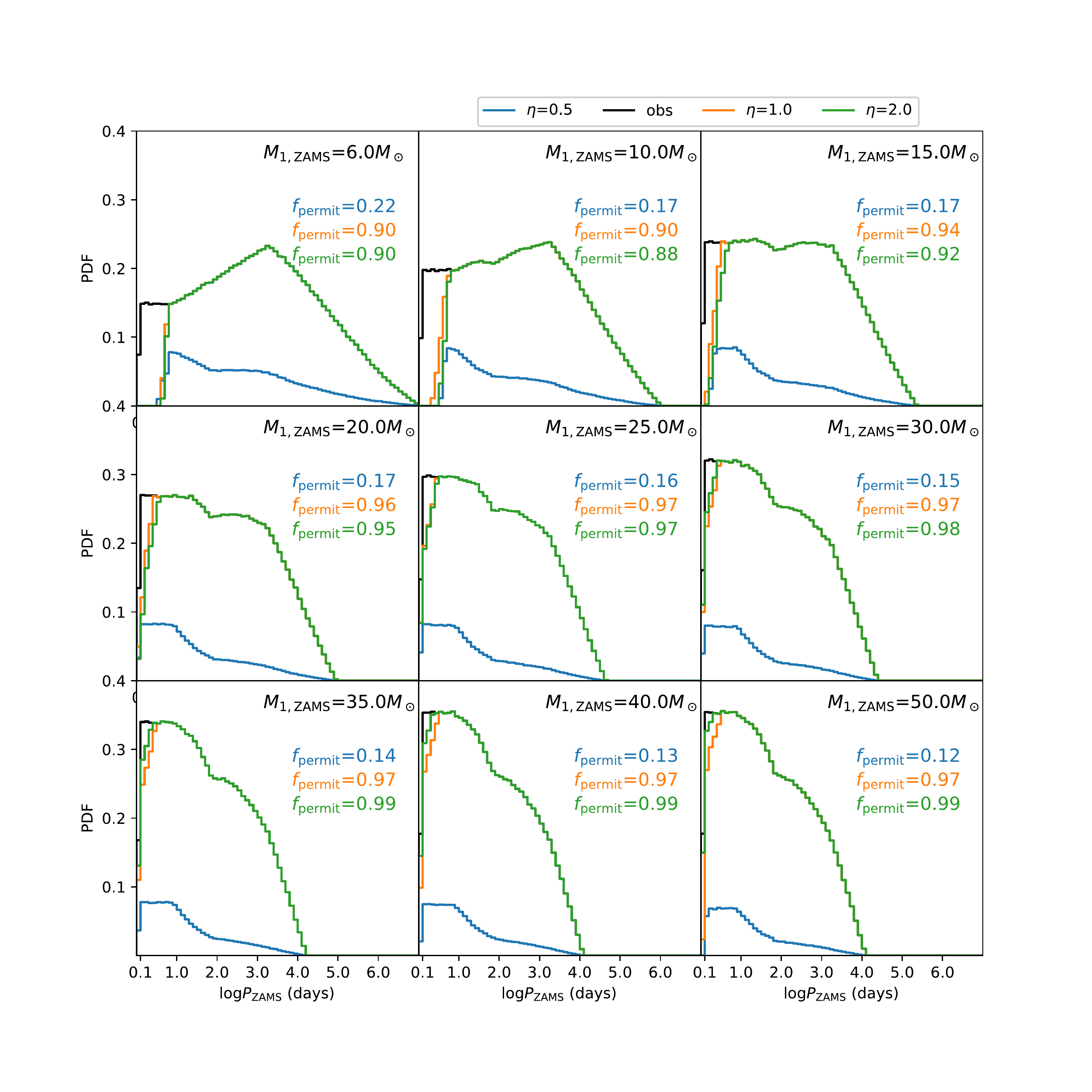}
    \caption{Each panel represent for a specific primary mass $M_{1,ZAMS}$ the observed probability distribution in orbital periods $\log P$ (black) for all $q$ at ZAMS. In a similar fashion are also plotted the probability distribution in $\log P$ of permitted systems for different accretion grids with $\eta$=0.5 (blue), $\eta$=1 (orange), and $\eta$=2 (green) respectively. For each value of $\eta$ we write the fraction of permitted to total number of systems for each primary mass and each curve representing an $\eta$-value has been normalized to the fraction of permitted systems $f_{\rm permit}$.}
    \label{fig:pdf_permit_logP_zams}
\end{figure*}

\begin{figure*}
    \centering
    \includegraphics[width=1.0\linewidth]{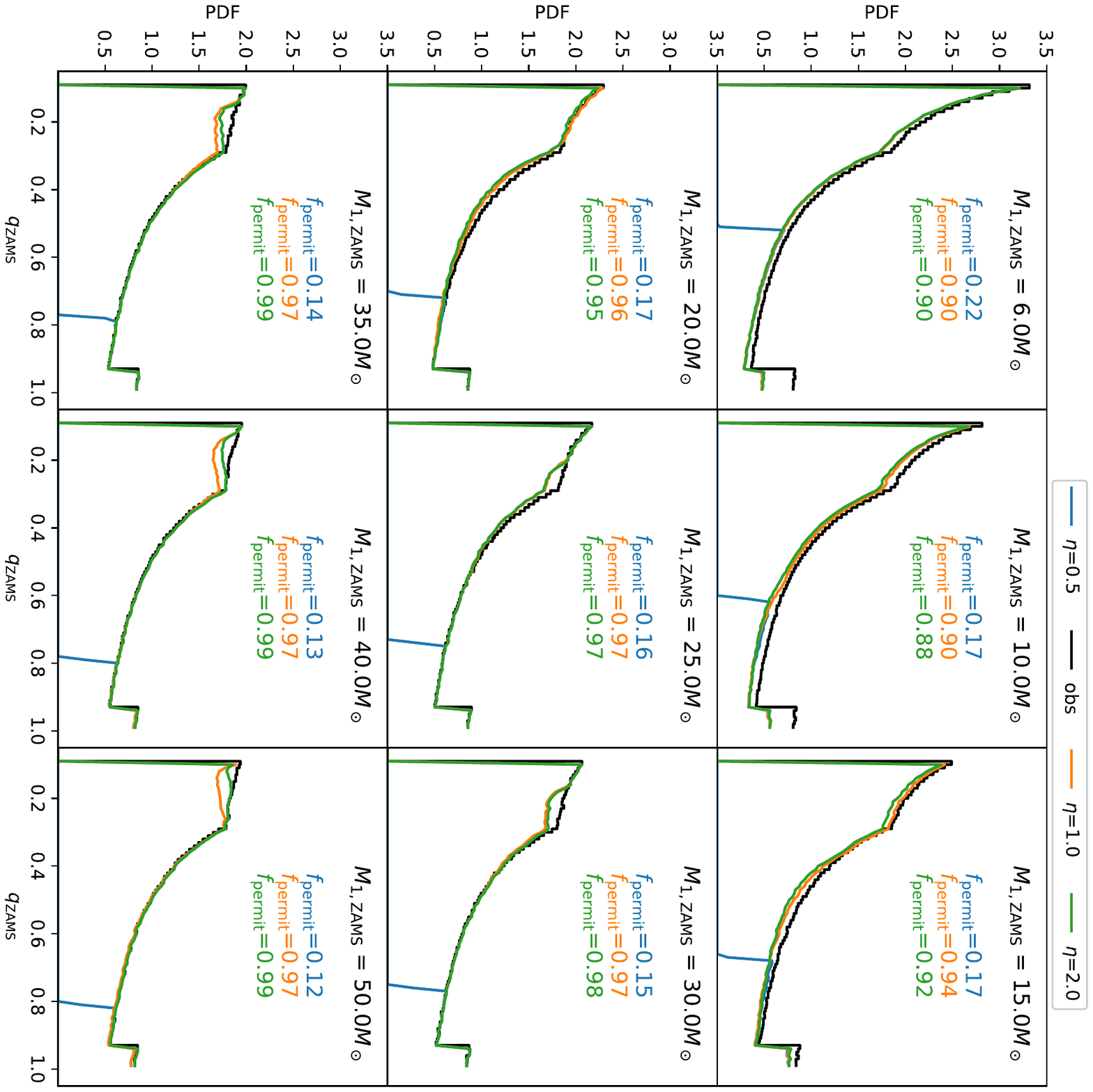}
    \caption{Same as Fig.~\ref{fig:pdf_permit_logP_zams} for the distribution in mass ratios $q$ for all $\log P$ and $e$ at ZAMS.}
    \label{fig:pdf_permit_q_zams}
\end{figure*}

In the two previous sections we have studied the case of systems containing a 40 M$_\odot$ primary mass at the ZAMS. Here 
we consider cases of different primary masses on the ZAMS between 6 and 50 M$_\odot$. For each case, we plots the probability distribution function for the ZAMS and initial orbital period and for the ZAMS and initial mass ratios (so plots similar to the side plots
in Figs. \ref{fig:q_logP_Mprim40eta2} and \ref{fig:pms_q_logP_Mprim40eta2}).
As done in previous sections, we divide the population of ZAMS binary systems between a forbidden and permitted region. Let us recall that the border drawn between these two regions is determined from our accretion models and is thus a limit entirely independent on the distribution of binary populations but a numerical output from our models.
An important question is what fraction of systems is permitted given some value of $\eta$? We seek an answer to this question in Figs. \ref{fig:pdf_permit_logP_zams} and \ref{fig:pdf_permit_q_zams}. 

Each panel in Fig. \ref{fig:pdf_permit_logP_zams} shows the $\log P$ distribution at ZAMS for all $q$ for the observed binary population (dashed black line), permitted population for $\eta=0.5$ (blue solid line), $\eta=1.0$ (orange solid line), and $\eta=2$ (green solid line), with each panel representing a primary mass $M_{1,ZAMS}$ = [6, 10, 15, 20, 25, 30, 35, 40, 50]\msun. Each curve representing an $\eta$-value has been normalized to the fraction of permitted systems $f_{\rm permit}$.

The observed population of binaries in $\log P$ display an interesting characteristic suggesting multiple formation mechanism and was noticed by \citet{tutukov1983}. The feature which suggest more than one formation process in play, is best observed for the orbital period distribution of the low mass range of primary stars. The orbital period distribution at primary masses of 6\msun{} is composed of function with a break. The break is seen at $\log P(\rm day)$= 3.5. The argument made by \citet{tutukov1983} is that the change in slope around the peak identifies a new regime of formation channel. Going from small to high primary masses, the pattern seems consistent across all primary masses however the break seems to move slightly inwards with increasing primary ZAMS mass. 

For all values of $\eta$, from $M_{1,ZAMS}$=6-15\msun{}, a population of very close binary systems are not reproduced by any of our models. The discrepancy is largest for $M_{1,ZAMS}$=6\msun{} and decreases for higher primary masses. From around 20\msun{} the discrepancy is very small. For very close binaries the discrepancy suggests they cannot have formed in a process of accretion only. A second general point is the fraction of systems reproduced by each model. Here it is seen that for $\eta=0.5$ the fraction of reproduced systems is significantly smaller compared to models of $\eta\geq$1 which are able to reproduce most systems.
The distribution for $\eta$=0.5 is biased towards high mass ratio systems since the mass ratio evolution is towards $q$=1 for any $\eta$<1. This is particularly shown in the tale of the $\eta$=0.5-distribution, which becomes smaller as one goes to higher primary masses. Hence, only close binaries are reproduced in this way.
The differences in the fraction of permitted systems if $\eta$=1 or $\eta$=2 are small. An interesting observation is the fraction of total systems these two $\eta$-values can reproduce. For $M_{1,ZAMS}$=6-20\msun{} $\eta$=1 is able to reproduce a marginal larger fraction of systems. At $M_{1,ZAMS}$=25\msun{} the permitted fraction are the same. For $M_{1,ZAMS}$>25\msun{}, models with $\eta$=2 reproduce a marginally larger fraction of systems than models with $\eta$=1. We also note, that the differences are always due to close binary systems.

\subsection{The mass ratios distributions at the ZAMS}

Each panel in Fig. \ref{fig:pdf_permit_q_zams} shows the $q$ probability distribution at ZAMS for all $\log P$ and $e$ for the binary population (black line) and the permitted population for $\eta=0.5$ (blue line), $\eta=1.0$ (orange line), and $\eta=2$ (green line), with each panel representing a primary mass $M_{1,ZAMS}$ = [6, 10, 15, 20, 25, 30, 35, 40, 50]\msun. The colored numbers in the figure shows the fraction of permitted to total systems reproduced by each model of different $\eta$-value. Each curve representing an $\eta$-value has been normalised to fraction of permitted system $f_{\rm permitted}$.

The overall characteristics of the observed mass ratio distribution is a two part power law with an additional excess for mass ratios $q$>0.95, known as the twin excess \citep[e.g.][]{lucy1979, tokovinin2000, krumholz2007, moe2017}. The range, which is an observational bias, has a lower limit of $q$=0.1. Observing binaries of $q$< 0.1 is very difficult as the primary star easily outshines the companion.

The permitted region for $\eta$=0.5 is limited to a range from $q$=0.5-1 for primaries of 6\msun{}. Moving to higher primary masses the lower limit of $q_{ZAMS}$ increases to $q$=0.8 for primaries of 50\msun. Hence, relative to $\eta\geq$1, the low fraction of permitted systems allowed/recovered for $\eta$=0.5 is due to the limit on the mass ratio evolution we inferred in eq. \eqref{eq:limqf} and Fig. \ref{fig:qi2qf}. This limit indicates, independent on the initial mass ratio, that increasing significantly the primary mass, would result in a final mass ratio of 1. As the majority of binaries observed are of low mass ratio, $\eta$ = 0.5 or smaller cannot recreate the range of observed binary systems in a pure accretion detached scenario.
We also note, that the distribution of $q_{\rm ZAMS}$ for twin binaries in the models for $\eta$=0.5 at primaries from 20\msun{} and up, decreases when $q_{\rm ZAMS}$ increases, 
while the actual distribution is flat. This indicates, that some binaries with mass ratios very close to 1 are not permitted at these high masses for $\eta$=0.5. The reason being that they are so close that Roche lobe overflow happens.

The permitted mass ratio distribution at ZAMS when $\eta$=1 or 2 (orange and green) are very similar in the fraction of systems they permit and in the shape of their curves. A common noticeable discrepancy of the two models against observations is at the low-end of the primary mass range, both models cannot reproduce the entire twin excess, indicating these twins did not form from accretion only or some post ZAMS evolution has taken place. For instance, would pre-MS mass transfer explain this as suggested in \citet[][]{krumholz2007}.
From primary masses of 15\msun{} and upwards the twin excess is well reproduced by both $\eta$-values and the difficulty in reproducing the distributions, is at low mass ratios, $q\simeq0.2$. The small dip at $q_{\rm ZAMS}$ around 0.2-0.3 is due to the secondary star overflowing its Roche lobe while undergoing the swelling phase.


\subsection{The protobinary orbital period and mass ratios distributions of detached binaries}

\begin{figure*}
    \centering
    \includegraphics[width=1.0\linewidth]{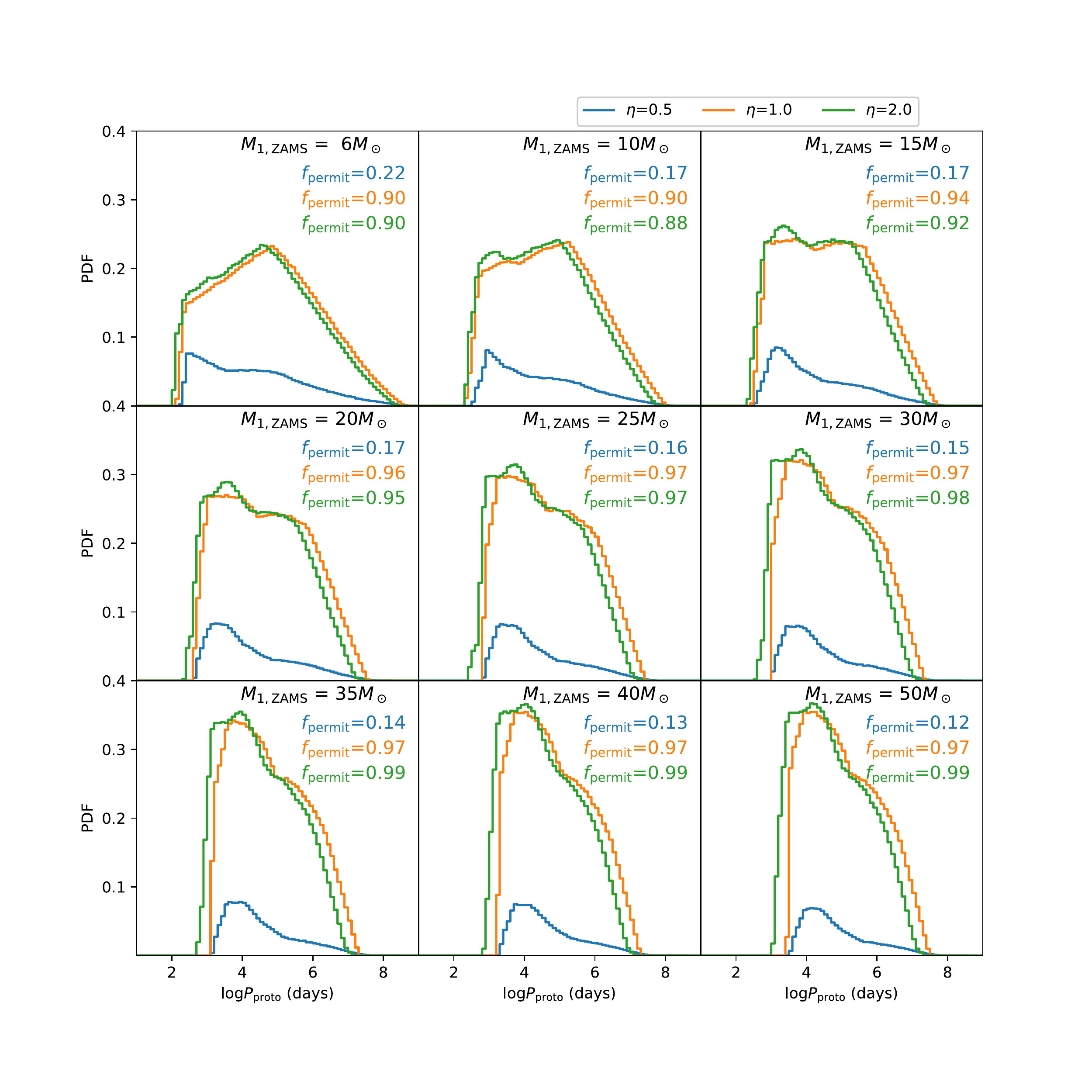}
    \caption{Same as Fig.~\ref{fig:pdf_permit_logP_zams} for the distribution in initial orbital periods 
    $\log P$ for all $q$ and $e$.}
    \label{fig:pdf_permit_logP_proto}
\end{figure*}

Let us first describe the protobinary orbital period distribution in Fig. \ref{fig:pdf_permit_logP_proto} by comparing it to Fig. \ref{fig:pdf_permit_logP_zams}, which is the ZAMS binary orbital period distribution. We see the effect of accretion is a shift from large to small orbital periods, which is expected from eqs. \eqref{eq:afai_disk} and \eqref{eq:pfpi_disk}, and are independent on the binary mass ratio changes.

The permitted distribution of protobinary mass ratios, for different $\eta$-values shows very different characteristics. For $\eta$=1 the protobinary distribution is the same as the ZAMS binary distribution, since $\eta$=1 fixes the mass ratio as a constant during accretion growth.

If $\eta$=0.5, we see a mass ratio distribution fairly similar to when $\eta=1$, but with a systematic difference. The systematic difference is that the $\eta=1$ distribution is elongated  towards lower mass ratios. For instance the mass ratio twin-excess part, is elongated to lower mass ratios, and keep doing so, as higher primary masses are reached. Had the grid of accretion sequence extended down to lower initial mass ratios than 0.1, a pile up of systems close to 0 would have been seen.
The protobinary mass ratio distribution for $\eta$=2 is different from the other two. For $\eta$=2 the accretion scenarios tends to decrease the mass ratio and therefore the initial mass ratio at onset of accretion is higher than the mass ratio at ZAMS. For primary masses of 6\msun{}, the range in mass ratios is from 0.4 to 1. As the primary mass increases, the minimum initial mass ratio moves to $\gtrsim$0.8 at 50\msun. Tis implies that initially the two protostars begins with nearly identical masses, and the subtle differences between the two masses leads to a \textit{run away}-effect that can decrease the mass ratio during accretion growth.

The range spanned by our inferred protobinary orbital period distribution at onset of accretion is very similar to the range spanned by pre-MS solar type binaries \citep{mathieu1994}.
If the observed range of solar type pre-MS binaries \citep[see][]{mathieu1994} is the range in which most of these binaries are found, it would suggest that this is the primary range at which such solar type binaries forms since they do not undergo strong accretion from an external reservoir. Consequently, had a pre-MS solar type binary formed within this range and subsequently undergone significant accretion growth, it would become massive and close. This suggests, that massive close binaries, would be the \textit{lucky} few low-mass pre-MS binaries that happens to undergo significant accreting growth. This is suggesting the formation of massive binary stars is no different from that of low-mass binaries. But the accretion growth phase is much more extensive in the case of high mass binaries.

\begin{figure*}
    \centering
    \includegraphics[width=1.0\linewidth]{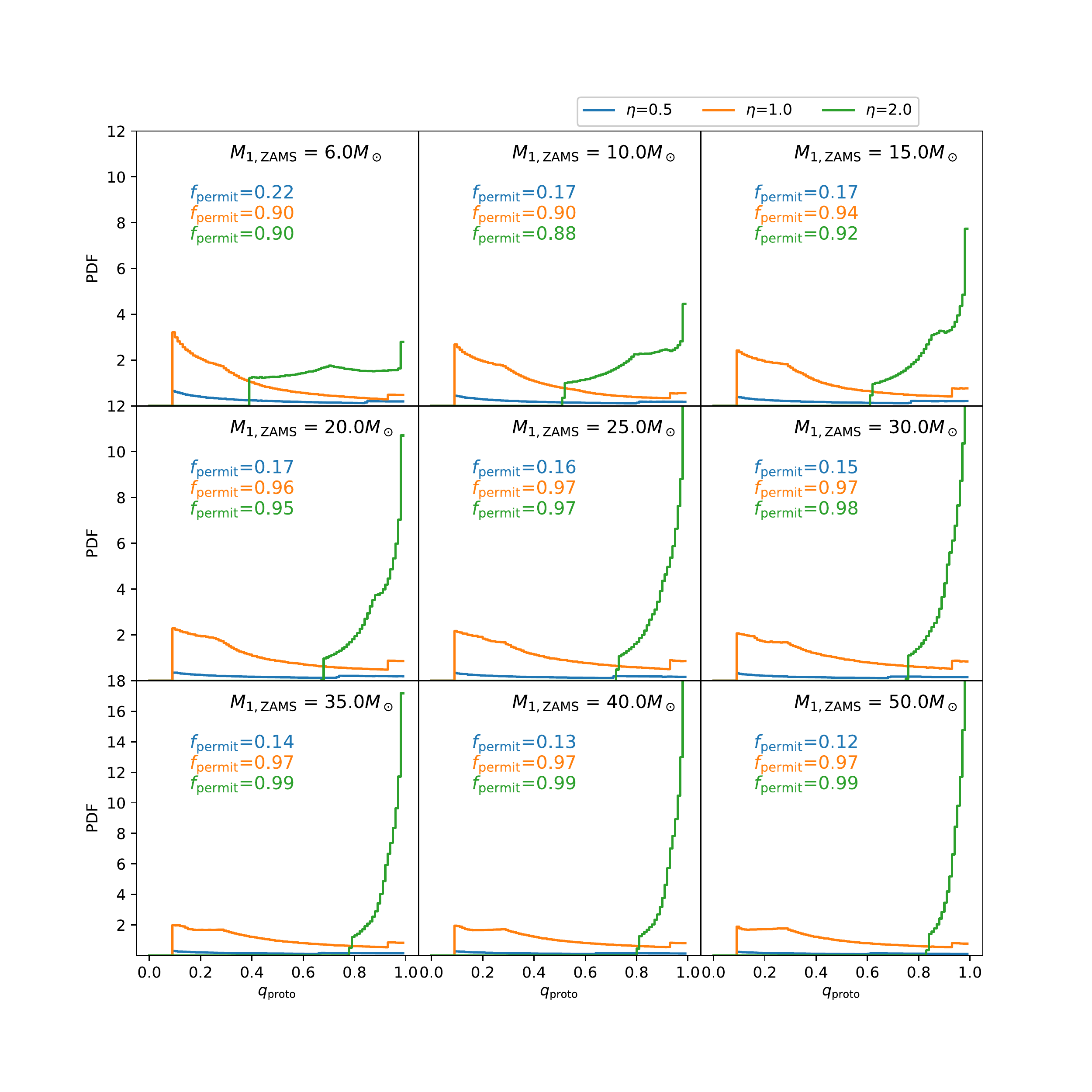}
    \caption{Same as Fig.~\ref{fig:pdf_permit_logP_zams} for the distribution in initial mass ratios 
    $q$ for all $\log P$ and $e$.}
    \label{fig:pdf_permit_q_proto}
\end{figure*}

\section{Discussion}\label{sec:discussion}

\subsection{Alternatives and missing physics}
The main uncertainties of this work stem from some potentially important physical processes or alternative models that have been ignored or not pursued here. These includes different accretion history profiles, stellar rotation, and tides.

From the perspective of stellar evolution, the key difference between the proposed massive star forming theories is one of accretion rates and secondly how these are achieved. \citet{McKee2002,mckee2003} and \citet{krumholz2007} assumes very high accretion rates, while the picture drawn by \citet{bonnell2001} is likely to propose a different accretion rate profile, in few cases, similar to a constant high accretion rate. We have produced a grid of pre-MS binary accretion sequences with constant accretion rate of $\dot{M} = 10^{-4}$\msun{} $yr^{-1}$ for $\eta=1$. Relative to our accretion history from the CH law, a constant accretion rate produces a slightly different radial profile with protostar mass. The primary difference is, that the swelling phase is less extended in mass at constant accretion rate. This happens because, the swelling phase in the CH law is followed by an increase in the stars luminosity, hence accretion rate, and this prolongs the swelling phase.

Including stellar rotation into the problem may have an impact on the maximum mass that a star can grow to. The stellar structure is affected by the angular momentum content of the protostar \citep{haemmerle2017}. As material is accreted onto the protostar, angular momenttum is also added, spinning it up. If allowed to continue, the protostar is spinning increasingly faster, potentially reaching a critical surface velocity whereby material cannot be accreted by the star any further, and material may even be lost from the protostellar surface \citep[e.g.][]{langer2012}.

The effect of tides is twofold. In eccentric orbits, tides can circuralise it. If the angular momentum content of material accreted by the protobinary, increases the orbital eccentricity, tidal circularisation can work against this effect. Secondly, tides work to synchronise the spin of the protostars to the orbital period. In cases where a protostar's spins at critical velocity, above the tidal synchronisation speed, the tidal force can decrease the spin to the orbital velocity. This potentially could allow the continuation of accretion at a lower rate. This however, requires, that the binary orbit is sufficiently small to allow a tidal force to be effective. It will be very relevant to model the accretion phase of protobinaries where such effects are taken into account.

\subsection{Comment on the \textit{angular momentum problem}}
A key open question in stellar astrophysics, concerning both the formation of stars and the later stages of stellar evolution \citep{langer2012}, is how angular momentum is transported. Stars form from cold low density gas which eventually becomes dense and hot, increasing the central density many orders of magnitude. This should lead to rapidly rotating stars due to angular momentum conservation. Further, when the stars evolve away from the main sequence they also increase in central compactness. This central contraction to increase the density would also increase the rotation of the star unless angular momentum can be dissipated by some mechanism. Somehow stars can avoid spinning critically due the contraction and this is known as the \textit{angular momentum problem}. It was suggested by \citet{larson1972am_problem} that all stars are formed in a binary- or higher order multiple system whereby the angular momentum of the infalling gas is given to the binary orbit rather than the stars them selves.

In the present paper we have, implicitly, explored this idea via disc accretion, in which binary stars form in initially wide orbits. The accretion growth then absorbs all the angular momentum into the binary orbit, which shrinks at a sufficiently low rate, that therefore allow the formation of very massive stars in close orbits.
The great unknown here is how valid our assumption that the accreted material always have the specific angular momentum of the accretor is. Accounting for rotation of each protostar would allow the protobinary to absorb slightly more angular momentum but only a small amount relative to the binary orbit.
Another binary effect not included here is that of a massive circumbinary disc. Dynamics and close encounters with other stars in a starforming complex may also play a key role in the transport of angular momentum.

\subsection{The value of $\eta$}

We have seen above, discussing various hydrodynamical simulations, that determining the value of $\eta$ is not easy and at the moment a large spread of values is obtained, mainly due to the different conditions considered and secondly due to differences in resolution between the models \citep{ochi2005}.
Despite this relatively low information content we presently have about the numerical value of $\eta$, we can make qualitative arguments about extreme values of $\eta$ by considering that any value of $\eta$ must allow the reproduction of the observed binary mass ratio range at ZAMS.  
We consider, qualitatively the two extreme cases of $\eta << 1$ and $\eta >> 1$. In the first case, the mass ratio will be driven to unity with very little accretion taking place.
This means that only high mass ratios systems can be formed in that way. These high mass ratios represent only a small fraction of the observed systems. Thus it suggests the most dominant star formation process of binaries are; capture events, fragmentation of the primary's circumstellar disc taking place late in the accretion phase, or that protostars become bound after they have undergone most of their accretion growth.

In the case $\eta >> 1$, because accretion drives the mass ratio towards zero, all protobinaries form nearly identical. In that scenario, protobinaries are formed as low mass binaries with a mass ratio above 0.8. They form either from fragmentation or from capture early in the accretion phase and accrete most if not all their mass as bound binary systems.
Our qualitatively estimated values of $\eta$ fall in the range from 1 to 5, and wee see that values of $\eta$ equal to 1 or 2 can both reproduce the observed distribution of ZAMS binaries quite.

\subsection{Roche lobe overflow at pre-main sequence}
\citet{krumholz2007} find that only the primary star is able to initiate a mass transfer. We find, on the contrary, that both the secondary star and the primary star can overflow their Roche lobe during accretion. 

Further, we find that those pre-MS binary accretion sequences, where the secondary protostar fills its Roche lobe, is during the swelling phase or when its accretion rate is very small, such that very low mass ratios are formed. We will avoid speculating about the outcome of such mass transfer sequences here, and leave this subject for a subsequent study..

If mass transfer happens at all, during the pre-MS accretion growth phase, remains an open question. \citet{krumholz2007} proposed a number of systems as candidates to potentially have experienced pre-MS mass transfer, e.g. WR20a \citep{bonanos2004,rauw2004} and a number of eclipsing OB binaries with orbital periods around 5 days in the SMC \citep{Harris2003,hilditch2005}. \citet{krumholz2007} assumed a much larger accretion rate relative to the CH accretion law used here, producing radially more extended stars. But this does not rule out their potential as pre-MS mass-transfer candidates. What makes an interpretation difficult here for these systems, is that the systems have evolved away from the ZAMS and it is therefore necessary to model the accretion phase of these systems and the main sequence of the system.
Another interesting system is the eclipsing massive binary LMC-SC1-105 which is currently undergoing stable mass transfer \citep{bonanos2009}.
A final system of great relevance is the massive overcontact binary VFTS352 which we already mentioned in Sect. \ref{sec:grids}.

A direct analysis of the above mentioned systems is not possible here, due to differences in metallicity or because the systems have evolved away from the main sequence. A future work will address the origin of these systems via detailed modelling of individual systems from pre-MS and until their present state. Doing so should help constrain the accretion history of these systems, as probably only a subset of theoretical birthlines, can explain the observational inferred parameters of these systems. Thus, potentially a wealth of information about massive star formation can be unlocked from studying the history of these systems. For instance it might be possible to test if one or more systems went through a pre-MS mass transfer phase or not.

\subsection{Observations of pre-MS binaries}
Observational evidence of the mass ratio and the orbital period distribution of pre-MS binaries is scarce. For massive stars, this is somewhat expected since we observe only a few massive binaries and rarely, if ever, while they are undergoing pre-MS accretion. The accretion phase is expected to be short lived and the short life time of massive stars in general makes them rare. For solar mass stars observations are more numerous. We also observe pre-MS solar type binaries. This raises the question, can we can link the observations of solar type pre-MS binaries to properties of star formation relevant at all mass ranges? What at least strikes as an interesting comparison is the range of orbital periods in which we observe pre-MS solar type binaries and the protobinary orbital period distribution at the onset of accretion we find. In \citet{mathieu1994} the orbital period of pre-MS solar type binaries was found to form a bimodal distribution with one component  in the range log p= 0-2 days and second dominant component in the range logP=3.5-7.5 days. Since such systems are expected to undergo little accretion growth, their observed orbital period is more or less the initial binary orbit. The binary systems in the upper orbital period range in \citet{mathieu1994} is similar to the orbital period range at onset of accretion for the inferred initial protobinaries we find and show in Fig. \ref{fig:pdf_permit_logP_proto}. It is noteworthy how we used a 1\msun{} primary protostar model and independently of the observations infer the same range. This could be a hint that even massive binary stars acquire their mass via accretion growth from some external reservoir as proposed by \citet{bonnell2001}, and that the earliest conception out of the gas phase is independent on the final mass of the star.

\subsection{Can we measure the total mass accretion rate of massive binary stars?}
The observations of S106 IR is a direct observation of a massive binary system at the pre-MS stage. A crucial open question in this respect is to estimate if and how the system is accreting, as might be indicated by the mysterious \textit{dark lane} connecting to the system \citep{comeron2018, Schneider2018}.
A potential path to enlighten this question is the beauty of the Kepler two body problem as we have explored it here, and specifically as expressed in the eq. \eqref{eq:pfpi_disk}, which relates the orbital period of the system to the total initial and final mass of the system assuming disc accretion. For instance, if we measure the orbital period change over a period of 10 yrs what would be the associated mass accretion rate causing the period change? This can be deduced from observations assuming we have a sufficient precise orbital period determination. To see this we differentiate eq. \eqref{eq:pfpi_disk} with respect to time, ($P_i$ and $M_i$ are of course constants), and we get
\begin{equation}\label{eq:orbit_decay}
    \dot{P}_{f} = -2 \frac{M_i}{M_f^2}\dot{M}_fP_i
\end{equation}
where $\dot{P}_{f}$ is in units of days yr$^{-1}$. Rewritting this equation with respect to relevant units and introduction a baseline observational time $\Delta T$ gives
\begin{equation}\label{eq:baseline}
    \dot{M} = \frac{1}{2}\left(1-\frac{P_f}{P_i}\right)\left(\frac{M_f}{M_{\odot}}\right)^2 \left(\frac{M_i}{M_{\odot}}\right)^{-1}\left(\frac{\Delta T}{yr}\right) \frac{M_{\odot}}{yr} \, ,
\end{equation}
which relates the baseline time and orbital period change to the mass accretion rate during the baseline time. Since, really, the accretion rate during the time interval is a mean value we simply write $\dot{M}$. If $P_f>P_i$ there has been a total mass loss, while $P_f<P_i$ indicate mass accretion. $P_f=P_i$ indicates no accretion.
The question now is, how accurate we need to measure the orbital period of S106IR to learn about the accretion rate into the system or to put an upper limit to the systems accretion rate given a $\Delta T = 10$yr baseline. S106IR has total mass $\sim$23\msun{} and is in an orbital period of 5~days, assuming a constant accretion rate we estimate the orbital period change pr. 10 year period. Massive binary stars like S106IR is expected to undergo fairly high accretion rates of $10^{-4}$\msun $yr^{-1}$ or higher assuming they are still in the accretion growth phase. 
From eq. \eqref{eq:baseline} therefore we can deduce that the orbital period must be determined to a precision of $10^{-3.5}$days or better to confirm such high accretion rates.

\section{Conclusion}\label{sec:conclusion}
In this work we have explored how binary star systems might become massive and close through a process of accretion that started from a low mass proto binary < 2\msun.
We have done this by formulating a family of analytic models to describe the change in orbital separation and mass ratio as the binary system undergoes accretion growth. The orbital separation evolution we described assumes a disc geometry under the assumption that the two binary stars accrete material that carry the specific orbital angular momentum of the accretring star. 
The mass accreted is shared between the two components according
to a law $\tfrac{\dot{M}_2}{\dot{M}_1}=q^{\eta}$ at all times. 

We combined our model of mass ratio and orbital separation evolution with the stellar evolution code MESA, and computed grids of pre-MS binary stars undergoing accretion growth, varying the initial orbital separation and mass ratio. We computed such grids for a range of $\eta$-values from 0.5 to 3. With these grids we explored the stellar structure and evolution in HR diagram of protobinaries undergoing accretion growth. 
From the grids we inferred the limits on how massive a binary system could become, given its initial conditions, before either of the binary stars would fill their Roche lobe, at which point our computations stopped. This yielded the result that accretion into binary stars during pre-MS can lead to both the primary star and the secondary star overflowing their Roche lobe.
A final step we took was to compare the inferred limits of our model grids for different values of $\eta$ with the observed population of binary stars from \citet{moe2017}.
From our efforts we note that:
\begin{itemize}
    \item Accretion via discs allows binary stars to accrete more mass before filling their Roche lobe compared to accretion from a spherical symmetric accretion process. This is similar to accretion growth of single stars, where disc accretion allows single stars to overcome the luminosity barrier (see eq. \eqref{eq:ai2af_spherical} and \eqref{eq:afai_disk}).
    \item The mass ratio evolution can be parameterised by $\eta$ such that $q^{\eta}=\tfrac{\dot{M}_2}{\dot{M}_1}$, and determines wether the mass ratio goes towards unity or zero, see eqs. \eqref{eq:qi2qf} and \eqref{eq:qf2qi} and Fig. \ref{fig:qi2qf}. Two special cases are $\eta=1$ in which case the mass ratio remains a constant and $\eta=0$ in which the binary total mass accretion is split equally by the two stars or $q=1$.
    If $\eta<1$ the mass ratio evolution is towards unity and this puts a limit on the mass ratio distribution that emerges after the accretion phase by introducing a minimum final/ZAMS mass ratio of binary stars forming from accretion even if the initial mass ratio is very small (see Fig. \ref{fig:eta_qi2qf}). 
    \item If $\eta>1$ the mass ratio evolution is towards zero and there is no upper or lower limit to the final mass ratio after end accretion growth.
    \item Estimating $\eta$ from hydrodynamic simulations of protobinaries is difficult and depends on the initial assumption of gas sound speeds and specific orbital angular momentum of the infalling gas, see Fig. \ref{fig:eta_q_estimates}.
    Using an analytic approach assuming Bondi-Hoyle accretion or Roche lobe radius we found that $\eta$ lies in the range 1-3.
    \item Both the primary and secondary protostars can overflow their Roche lobe and whether the former or the latter occurs is strongly correlated with the swelling phase of either star, see Figs. \ref{fig:grid_min_aiM1zams_eta1p0} and \ref{fig:grid_min_aiM1zams_eta2p0}.
    \item The minimum orbital period at ZAMS decreases with increasing primary mass and increasing mass ratio. This allows for the formation via accretion growth of chemically homogeneously stars in close binaries (see Figs. \ref{fig:minLogP_0.5} and \ref{fig:minLogP_2.0}). Systems like VFTS352 may have formed from accretion growth and ending as a contact system.  It also seems that this result is somewhat insensitive to the value of $\eta$. 
    \item For $\eta=1$ or $\eta=2$, the mass ratio and orbital period distribution of most binary systems, 88 to 99\%, for a primary star mass range from 6 to 50 \msun{} could be reproduced (see Figs. \ref{fig:pdf_permit_logP_zams} and \ref{fig:pdf_permit_q_zams}). The massive pre-MS binary S106IR \citep{comeron2018} can also be produced from the accretion channel if $\eta$ is 1 or 2.
    \item From the ZAMS distribution of binary stars we infer the protobinary mass ratio distribution at the onset of accretion and find the orbital period range to coincide with pre-MS solar type stars \citep{mathieu1994}, that likely have not accreted significantly. See Figs. \ref{fig:pdf_permit_logP_proto} and \ref{fig:pdf_permit_q_proto}.
    \item The picture of massive binary stars forming at small mass and subsequent accretion growth is a likely explanation for the distribution of the observed range of massive binary stars, and could allow for the formation of progenitors of coalescing binary black holes.
\end{itemize}
There are still many open questions on how binary systems form and we have not here been able to account for all possible relevant physics. In brief we mention; eccentric orbits, tidal forces, and dynamic effects, which will affect the conclusion of this paper and we look forward to continuing this effort exploring more physics.

\begin{acknowledgements}
    We thank Smadar Naoz and Kaitlin Kratter for fruitful discussions during the makings of this work. We also thank Max Moe for sharing his script to sample binary populations from his joint probability distribution that was used in this work.
    This project was supported by the Swiss National Science Foundation project number 200020-172505. The computations were performed at University of Geneva on the Baobab cluster.
    We at DARK are grateful for support from the DNRF (Niels Bohr Professorship Program), the Carlsberg Foundation and the VILLUM FONDEN (project number 16599). MS is grateful to the hospitality offered by DARK during the making of this paper.
    This project used extensively Python, numpy, matplotlib and Astropy \citep{hunter2007, astropy}.
\end{acknowledgements}

\bibliographystyle{apalike}

\begin{appendix}
\section{Estimates of $\eta$ from \citet{bate1997gas}, \citet{ochi2005}, and \citet{young2015b}}\label{appendix:eta}
Figure 4 in \citet{bate1997gas} shows the relative accretion rate onto $M_1$ and $M_2$ in their simulations. These are thus corresponding to $f_1$ and $f_2$ introduced in sect. \ref{sec:mass_ratio_evo}. Finding $\eta$ is done by reading $f_1$ in the figure and the condition $1= f_1 + f_2$.
We note that the models of \citet{bate1997ballistic} shows a peculiar behaviour for q=1.0 where we would expect $\eta$=1 and mass accretion fractions of 0.5 onto each star. This is also the case for $j_{\inf} \leq 0.8$ but changes for higher initial specific angular momenta.
\begin{table}[]
    \centering
    \caption{Estimates of $\eta$ based on \citet{bate1997gas} which are read of from their Fig. 4.}
    \label{tab:etaBB1997}
    \begin{tabular}{cccc}
    \hline
    \hline
    $q$ & $j_{\rm inf}$ & $f_1$  & $\eta$ \\
    0.1 & 0.6 & 0.975 & 1.59 \\
0.1 & 0.8 & 0.55 & 0.09 \\
0.1 & 1.0 & 0.1 & -0.95 \\
0.1 & 1.2 & 0.05 & -1.28 \\
0.1 & 1.4 & 0.1 & -0.95 \\
0.2 & 0.4 & 0.95 & 1.83 \\
0.2 & 0.6 & 0.725 & 0.60 \\
0.2 & 0.8 & 0.675 & 0.45 \\
0.2 & 1.0 & 0.125 & -1.21 \\
0.2 & 1.2 & 0.05 & -1.83 \\
0.4 & 0.0 & 0.95 & 3.21 \\
0.4 & 0.2 & 0.85 & 1.89 \\
0.4 & 0.4 & 0.7 & 0.92 \\
0.4 & 0.6 & 0.6 & 0.44 \\
0.4 & 0.8 & 0.35 & -0.68 \\
0.4 & 1.0 & 0.2 & -1.51 \\
0.4 & 1.2 & 0.1 & -2.40 \\
0.4 & 1.4 & 0.05 & -3.21 \\
0.6 & 0.0 & 0.825 & 3.04 \\
0.6 & 0.2 & 0.675 & 1.43 \\
0.6 & 0.4 & 0.6 & 0.79 \\
0.6 & 0.6 & 0.55 & 0.39 \\
0.6 & 0.8 & 0.35 & -1.21 \\
0.6 & 1.0 & 0.25 & -2.15 \\
0.6 & 1.2 & 0.15 & -3.40 \\
0.6 & 1.4 & 0.075 & -4.92 \\
0.8 & 0.0 & 0.65 & 2.77 \\
0.8 & 0.2 & 0.55 & 0.90 \\
0.8 & 0.4 & 0.525 & 0.45 \\
0.8 & 0.6 & 0.5 & -0.00 \\
0.8 & 0.8 & 0.35 & -2.77 \\
0.8 & 1.0 & 0.3 & -3.80 \\
0.8 & 1.2 & 0.2 & -6.21 \\
0.8 & 1.4 & 0.1 & -9.85 \\
0.8 & 1.6 & 0.05 & -13.20 \\
1.0 & 0.0 & 0.5 & - \\
1.0 & 0.2 & 0.5 & - \\
1.0 & 0.4 & 0.5 & - \\
1.0 & 0.6 & 0.5 & - \\
1.0 & 0.8 & 0.35 & - \\
1.0 & 1.0 & 0.4 & - \\
1.0 & 1.2 & 0.325 & - \\
1.0 & 1.4 & 0.15 & - \\
1.0 & 1.6 & 0.5 & - \\
1.0 & 1.8 & 0.0 & - \\
1.0 & 2.0 & 0.0 & - \\
    \hline
    \end{tabular}
\end{table}

To estimate $\eta$ from \citet{ochi2005} we read their table 1. We selected only those models with sound speeds $c_s = 0.25$ as these where the most representative. Like for \citet{bate1997gas}, the models of \citet{ochi2005} the mass accretion fraction when $q=1$ is not following an equal mass accretion onto each protostar which one would otherwise expect.

\begin{table}[]
    \centering
    \caption{Estimates of $\eta$ based on \citet{ochi2005} which are taken from their table 1.}
    \label{tab:etaOchi2005}
    \begin{tabular}{cccc}
    \hline
    \hline
    $q$ & $j_{\rm inf}$ & $\frac{\dot{M}_2}{\dot{M}_1}$ & $\eta$ \\
    0.2 & 1.2 & 0.85 & 0.10 \\
0.2 & 1.3 & 0.54 & 0.39 \\
0.3 & 1.2 & 0.65 & 0.35 \\
0.3 & 1.3 & 0.45 & 0.66 \\
0.4 & 1.3 & 1.15 & -0.15 \\
0.4 & 1.4 & 1.25 & -0.24 \\
0.4 & 1.5 & 0.52 & 0.72 \\
0.4 & 1.6 & 0.41 & 0.97 \\
0.4 & 1.7 & 0.33 & 1.20 \\
0.4 & 1.8 & 0.45 & 0.88 \\
0.4 & 1.9 & 0.53 & 0.68 \\
0.5 & 1.3 & 0.50 & 1.00 \\
0.5 & 1.4 & 0.55 & 0.86 \\
0.5 & 1.5 & 0.53 & 0.90 \\
0.5 & 1.6 & 0.40 & 1.32 \\
0.5 & 1.7 & 0.35 & 1.53 \\
0.5 & 1.8 & 0.37 & 1.42 \\
0.5 & 1.9 & 0.39 & 1.34 \\
0.6 & 1.3 & 0.50 & 1.37 \\
0.6 & 1.4 & 0.50 & 1.37 \\
0.6 & 1.5 & 0.56 & 1.15 \\
0.6 & 1.6 & 0.54 & 1.20 \\
0.6 & 1.6 & 1.19 & -0.34 \\
0.6 & 1.6 & 0.82 & 0.39 \\
0.6 & 1.6 & 0.55 & 1.16 \\
0.6 & 1.7 & 0.37 & 1.97 \\
0.6 & 1.8 & 0.37 & 1.95 \\
0.6 & 1.9 & 0.35 & 2.07 \\
0.7 & 1.4 & 0.59 & 1.47 \\
0.7 & 1.5 & 0.41 & 2.48 \\
0.7 & 1.6 & 0.20 & 4.53 \\
0.7 & 1.7 & 0.36 & 2.89 \\
0.7 & 1.8 & 0.29 & 3.44 \\
0.7 & 1.9 & 0.43 & 2.38 \\
0.8 & 1.5 & 0.30 & 5.34 \\
0.8 & 1.6 & 0.31 & 5.21 \\
0.8 & 1.7 & 0.34 & 4.89 \\
0.8 & 1.8 & 0.36 & 4.60 \\
0.8 & 1.9 & 0.55 & 2.71 \\
0.9 & 1.5 & 1.01 & -0.08 \\
0.9 & 1.6 & 0.67 & 3.78 \\
0.9 & 1.7 & 0.98 & 0.19 \\
0.9 & 1.8 & 0.49 & 6.81 \\
0.9 & 1.9 & 0.78 & 2.42 \\
1.0 & 1.5 & 0.66 & - \\
1.0 & 1.6 & 1.01 & - \\
1.0 & 1.7 & 0.53 & - \\
1.0 & 1.8 & 0.94 & - \\
1.0 & 1.9 & 1.10 & - \\
    \hline
    \end{tabular}
\end{table}

In \citet{young2015b} the estimate of $\eta$ is done by reading a value for $\lambda$, equal to our $f_1$, from fig. 2, by trying to put a straight line through each of their models. Fig. 2 in \citet{young2015b} is for cold accretion models. Their Fig. 3 describes models of hot accretion and here the mass accretion fraction onto each is fluctuating such that no easy mean accretion rate can be read of from the figure, except for the case of $q$=0.9 which seems close to but, slightly less than $f_1$=0.5 on average. We note \citeauthor{young2015b} have no models for q=1 but assume $f_1$=0.5, hence $\eta$=1 for these models. Considering the effect in $\eta$ that their models produce
\begin{table}[]
    \centering
    \caption{Estimates of $\eta$ based on \citet{young2015b} read from their fig. 2.}
    \label{tab:etayoungclarke2005}
    \begin{tabular}{cccc}
    \hline
    \hline
    $q$ & $j_{\rm inf}$ & $f_1$ & $\eta$ \\
    0.1 & 1.2 & 0.06 & -1.19 \\
0.2 & 1.2 & 0.12 & -1.24 \\
0.3 & 1.2 & 0.15 & -1.44 \\
0.4 & 1.2 & 0.17 & -1.73 \\
0.5 & 1.2 & 0.25 & -1.58 \\
0.6 & 1.2 & 0.3 & -1.66 \\
0.7 & 1.2 & 0.37 & -1.49 \\
0.8 & 1.2 & 0.42 & -1.45 \\
0.9 & 1.2 & 0.45 & -1.90 \\
    \hline
    \end{tabular}
\end{table}

\end{appendix}

\end{document}